\documentclass[journal]{IEEEtran}

\usepackage{cite}

\ifCLASSINFOpdf
   \usepackage[pdftex]{graphicx}
   
\else
\fi
\usepackage{amsmath}
\usepackage{bm}
\usepackage{amssymb}
\DeclareMathOperator*{\argmin}{argmin}

\usepackage{algorithmic}

\usepackage{array}
\renewcommand{\tablename}{TABLE~}

\newtheorem{corollary}{\bf Corollary}
\newtheorem{theorem}{\bf Theorem}

\newtheorem{lemma}{\bf Lemma}

\newtheorem{definition}{\bf Definition}

\begin{document}

%
%
%
%

\title{The Random Frequency Diverse Array: A New Antenna Structure for Uncoupled Direction-Range Indication in Active Sensing}
\author{Yimin~Liu,~\IEEEmembership{Member,~IEEE},
Hang Ruan,
Lei Wang,
and Arye~Nehorai,~\IEEEmembership{Fellow,~IEEE}       
\thanks{Y. Liu, H. Ruan, and L. Wang are with the Department
of Electronic Engineering, Tsinghua University,  Beijing, 100084, China.
e-mail: yiminliu@tsinghua.edu.cn, ruanh15, lei-wang14@mails.tsinghua.edu.cn.}
\thanks{A. Nehorai is with the Department of Electrical and Systems Engineering,
Washington University, St. Louis, MO 63130 USA. e-mail: nehorai@ese.wustl.edu.}
\thanks{The work of Y. Liu was supported by the National Natural Science Foundation of China (Grant No. 61571260 and 61201356). The work of A. Nehorai was supported by the AFSOR (Grant No. FA9550-11-1-0210). Corresponding e-mail: yiminliu@tsinghua.edu.cn.}
\thanks{The results of this paper were presented in part at the International Conference on Acoustics, Speech, and Signal Processing (ICASSP), Shanghai, China, March 2016 \cite{yimin2016ICASSP}.}}

\maketitle

\begin{abstract}
In this paper, we propose a new type of array antenna, termed the Random Frequency Diverse Array (RFDA), for an uncoupled indication of target direction and range with low system complexity. In RFDA, each array element has a narrow bandwidth and a randomly assigned carrier frequency. The beampattern of the array is shown to be stochastic but thumbtack-like, and its stochastic characteristics, such as the mean, variance, and asymptotic distribution are derived analytically. Based on these two features, we propose two kinds of algorithms for signal processing. One is matched filtering, due to the beampattern's good characteristics. The other is compressive sensing, because the new approach can be regarded as a sparse and random sampling of target information in the spatial-frequency domain. Fundamental limits, such as the Cram\'{e}r-Rao bound and the observing matrix's mutual coherence, are provided as performance guarantees of the new array structure. The features and performances of RFDA are verified with numerical results. 
\end{abstract}

\begin{IEEEkeywords}
Array antenna, frequency diverse, compressive sensing, Cram\'{e}r-Rao bound.
\end{IEEEkeywords}

%
\IEEEpeerreviewmaketitle

\section{Introduction}

\IEEEPARstart{T}{he} array antenna is very popular in active sensing techniques, such as radar, sonar, and ultrasonic detection\cite{van2004detection}, for reasons such as flexibility in beam steering over a range of directions, and its efficient system resource management \cite{merrill2001introduction}. In most conventional array structures, only the directions of the beampattern can be formed. But the range (or delay) features are also usually important, especially in target indication or active imaging applications \cite{merrill2001introduction}. 

\begin{figure}[!t]
\centering
\includegraphics[width=.5\textwidth]{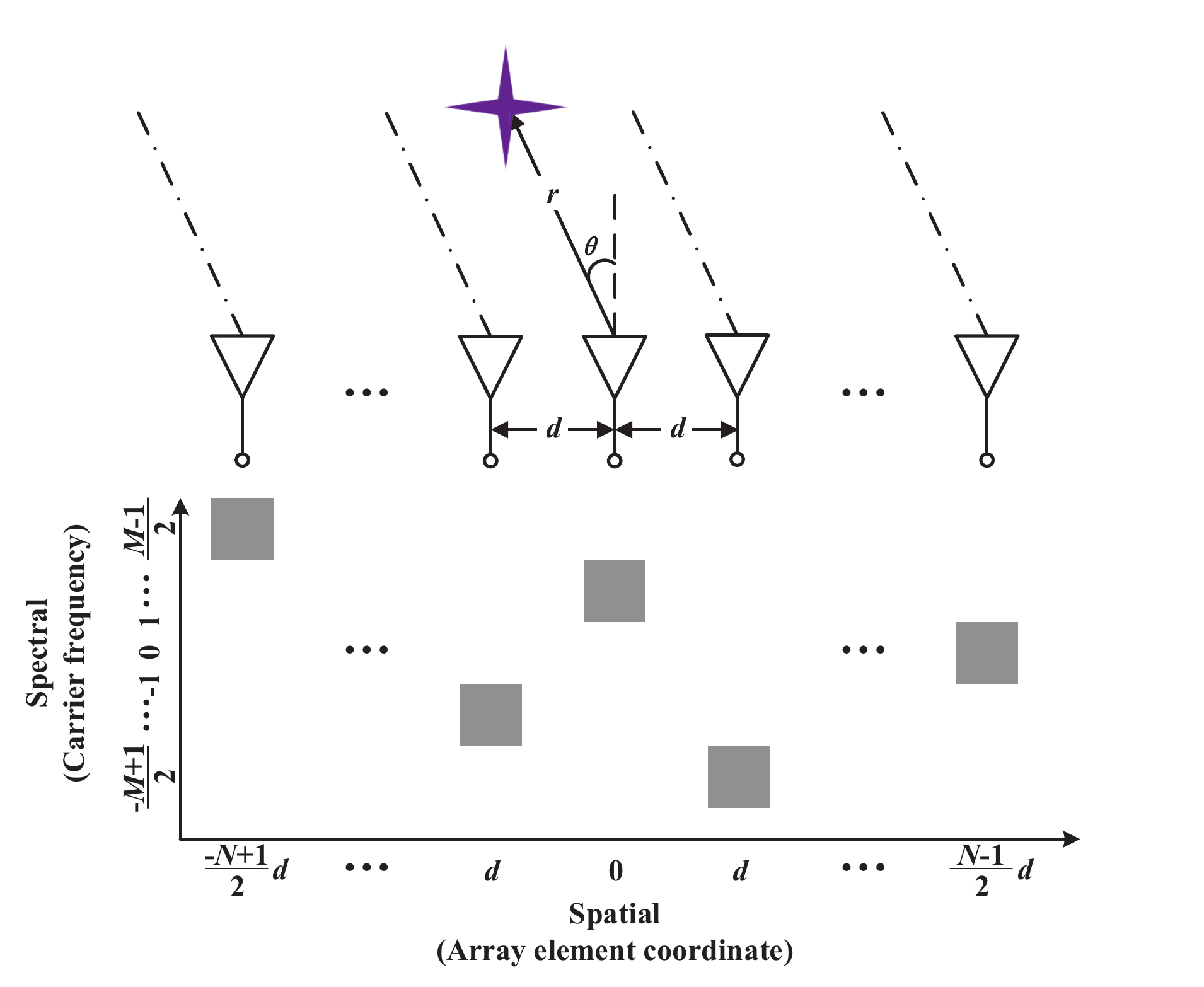}
\caption{System sketch of  RFDA.}
\label{Fig:SystemSketch}
\end{figure}

A promising new array structure, named the Frequency Diverse Array (FDA)\cite{Antonik2006Frequency} was proposed in 2006, and has attracted more and more attention \cite{wang2016overview} in recent years. The FDA can produce a beampattern with controllable direction and range by linearly shifting the carrier frequencies across the array\footnote{For abbreviation, we named it as Linear Frequency Diverse Array (LFDA) in the following discussion.}. This new feature provides many advantages in active sensing, such as low probability of interception and jamming resistance \cite{xu2015space}, resolution enhancement in Synthetic Aperture Radar (SAR) imaging \cite{farooq2008exploiting}, and ambiguous range clutter suppression in space-time adaptive processing \cite{xu2015range}.

However, as discussed in \cite{higgins2009analysis} and \cite{huang2010frequency}, the direction and range of a FDA's beampattern are coupled, which means there might be a group of direction-range pairs that match well with the echo from a point target, hence introducing aliases in target indication. To overcome this problem, several methods have been introduced recently. In \cite{wang2014range}, one straightforward method, named ``double-pulse", can effectively remove the coupling by successively transmitting two pulses. The direction and range indication are accomplished separately in each pulse, but the data rate drops to half.

The subarray, introduced by \cite{wangoptimal} and \cite{wang2014transmit}, is an alternative. In this approach, the FDA is divided into serval subarrays, each with a different carrier frequency increment. Uncoupled direction and range can be estimated, and the corresponding accuracy can be improved by optimizing the coordinates of the array elements and the frequency increments. However, only a single-target scenario is considered in these works. 

Multi-Input-Multi-Output (MIMO) radar \cite{Li2008MIMO} is a powerful technique with notable improvements, including the antenna aperture, spatial diversity, degrees of freedom. In \cite{sammartino2013frequency}, a new structure named Frequency Diverse MIMO (FD-MIMO) was proposed by combining MIMO and FDA. In FD-MIMO, each receiving element can demodulate all the signals from every transmitting element, each of which has a different carrier frequency. This design enables forming uncoupled direction-range beampattern by combining all the baseband samples, and dramatically increases the system's degrees of freedom. The FD-MIMO approach locates both single and multiple targets very well. However, the system complexity is also seriously increased due to the need for multiple receiver channels in each receiving element.

In this work, we regard the acquiring of a target's direction-range information by FDA as a sampling procedure in both the spatial and frequency domains. Then we propose a new system structure for a FDA, named Random Frequency Diverse Array (RFDA), which randomly assigns the carrier frequency of each array element. Inspired by the basic ideas of compressive sensing, RFDA can get the target information from both the spatial and frequency domains in a single glance. Through two-dimensional (2D) random sparse sampling, the new structure provides a thumbtack-like 2D beampattern, and can simultaneously achieve good resolution and estimation accuracy for both direction and range. The mutual coherence \cite{ben2010coherence} of RFDA's observing matrix is also quite small, which implies the ability of uncoupled target location in both single and multiple target scenarios. Moreover, each element in RFDA needs only one receiver channel, hence the above advantages can be attained without increasing the system complexity. The main contributions of this paper are as follows.

\begin{enumerate}
\item
We propose a new array structure, named RFDA, which randomly assigns the carrier frequencies of elements in a Uniform Linear Array (ULA). The new structure is shown to have the ability to simultaneously indicate target's direction and range without coupling.

\item
We derive the analytic form of RFDA's beampattern, and provide its stochastic characteristics, such as the mean, variance, and asymptotic distribution, to reveal the resolution and sidelobe level of this new array structure.

\item
We propose two signal processing methods for RFDA, matched filtering and compressive sensing, to detect the targets and indicate their locations. Matched filtering has the highest output signal-to-noise-ratio (SNR), while compressive sensing is more suitable for multi-target scenarios. Moreover, we give an approximately equivalent approach for matched filtering, to reduce the computation load.

\item
We derive the performance limits for RFDA, including the Cram\'{e}r-Rao Bound (CRB) and mutual coherence, to quantify its direction/range estimation and target indication abilities. 
\end{enumerate}

The rest of this paper is organized as follows. Section \ref{sec:SignalModel} presents a system sketch and constructs the signal model. In Section \ref{Sec:BeamPattern}, the beampattern and stochastic characteristics of RFDA are derived. In Section \ref{sec:SignalProcessing}, we introduce two signal processing methods, matched filtering, and compressive sensing. Section \ref{Sec:PerformanceAnalysis} investigates the performance limits of RFDA, such as the CRB and mutual coherence. Finally, numerical illustrations are given in Section \ref{sec:Results},  and conclusions are drawn in the last section.

\textit{Notations}: $\|\cdot\|_0$ is the non-zero entry number of an argument; $\|\cdot\|_{2,0}$ is the number of rows with non-zero $l_2$ norms in an argument; $\mathbb{E}_{\dagger}\{\cdot\}$ and $\mathbb{V}_{\dagger}\{\cdot\}$ are the expectation and variance of arguments with respect to (w.r.t.) the random vector (or variable) $\dagger$, respectively; $[\cdot]_{i,n}$ is the $i$th row, $j$th column of an argument; $[\cdot]_n$ is the $n$th element (or column) of an argument, while the argument is a vector (or matrix); $[\cdot]^*$, $[\cdot]^T$, and $[\cdot]^H$ are the conjugation, transpose, and conjugate transpose of arguments, respectively; $\mathbf{A}\succeq\mathbf{B}$ implies that $\mathbf{A}-\mathbf{B}$ is positive semi-definite; $\Re\{\cdot\}$ and $\Im\{\cdot\}$ are the real and imaginary part of an argument; The operator $\otimes$ means the Kronecker product, while $\odot$ means the Hadamard product.

\section{System Sketch and Signal Model}
\label{sec:SignalModel}
In this section, we will introduce a system sketch of the random frequency diverse array radar, and then propose the signal model of this new kind of array.

A brief system sketch of the RFDA is shown in Fig. \ref{Fig:SystemSketch}. There are $N$ elements in a linear array, with a constant inter-element distance $d$. These array elements are located symmetrically along the $x$-axis, and the coordinate of the $n$th one is
\begin{equation}
\label{Eq:AntennaLocation}
x_n=\left(-\frac{N-1}{2}+n\right)d, n=0,1,\dots, N-1.
\end{equation} 

Each element is connected to a narrow band transceiver. The transmitted waveform of each element is monotone sinusoid, but the carrier frequencies of the different elements are randomly assigned. The carrier frequency of the $n$th element is
\begin{equation}
\label{Eq:RFFrequency}
f_n=f_c+m_n\Delta{f},
\end{equation}
where $f_c$ is the center frequency, $\Delta{f}$ is the frequency increment, and $m_n$ is a random variable. In this work, all the $m_n$s are chosen as \textit{i.i.d.} random variables, and can be arranged into a random vector, $\mathbf{m}=[m_0, m_1, \dots, m_{N-1}]^T$.

In this paper, $g(m_n)$, which is assumed to be an even function, is defined as the probability density function (PDF) of $m_n$. And in the following investigation, we will take three common PDFs as examples:
\begin {enumerate}
\item
Gaussian: $g(m_n)={1}/{\sqrt{2\pi\sigma^2}}\cdot e^{-\frac{m_n^2}{2\sigma^2}}$, where $\sigma^2$ is the variance. The larger the $\sigma^2$ is, the wider the bandwidth of RFDA becomes.

\item
Continuous uniform: $g(m_n)={1}/{M}$, where $M$ is a positive real number, and $m_n\in[-M/2,M/2]$. The total bandwidth of RFDA is $M\Delta{f}$.

\item
Discrete uniform: $g(m_n)={1}/{M}$, where $N$ is a positive integer, and $m_n\in\left\{-({M-1})/{2}, -{(M-1)}/{2}+1,\dots,({M-1})/{2} \right\}$. The total bandwidth of RFDA is also $M\Delta{f}$. 
\end{enumerate}

Besides the above three, there can be other kinds of distributions for the frequency assignments of RFDA. With these definitions, the transmitted waveform of the $n$th element is
\begin{equation}
\label{Eq:TransSignal}
s_n(t)=e^{j2\pi(f_c+m_n\Delta{f})t}.
\end{equation}

A simple diagram of RFDA's waveform is illustrated in the lower part of Fig. \ref{Fig:SystemSketch}. The blue blocks illustrate the waveform distribution in the spatial-frequency domain. In this antenna, the original point of the $x$-axis, $O$, is selected as the phase center of the array. If there exists an ideal unit point target with direction $\theta$ and range $r$, then with the far-field assumption, the received radio-frequency (RF) echo of the $n$th element is
\begin{eqnarray}
\label{Eq:ReceSignal}
r_n(t;\theta,r)&=&s_n(t-2\frac{r+x_n\sin\theta}{c})\nonumber\\
&=&e^{j2\pi(f_c+m_n\Delta{f})\left(t-2\frac{r+x_n\sin\theta}{c}\right)},
\end{eqnarray}
where $c$ is the wave propagation speed. In each receiver, the RF echo is demodulated with its transmitting carrier frequency, so the baseband signal of the $n$th element is
\begin{eqnarray}
\label{Eq:BasebandScalar}
b_n(\theta,r)&=&e^{-j\frac{4\pi}{c}(f_c+m_n\Delta{f})\left[r+\left(-\frac{N-1}{2}+n\right)d\sin\theta\right]}\nonumber\\
&\approx&e^{-j\frac{4\pi}{c}\left[f_cr+(n-\frac{N-1}{2})f_cd\sin\theta+m_n\Delta{f}r\right]}.
\end{eqnarray}
As usually supposed in previous research works \cite{sammartino2013frequency}, the frequency increment $\Delta{f}$ is far less than $f_c$, which makes the approximation in  (\ref{Eq:BasebandScalar}) valid.

The baseband samples $b_n(\theta,r)$ of all elements ($n=0,1,\dots, N-1$) can be arranged in order to formulate a direction-range dependent steering vector:
\begin{eqnarray}
\label{Eq:SteeringVector}
\mathbf{b}(\theta,r)&=&[b_0(\theta,r), b_1(\theta,r), \dots, b_{N-1}(\theta,r)]^T\nonumber\\
&=&\mathbf{b}_D(\theta)\odot\mathbf{b}_R(r),
\end{eqnarray}
where
\begin{eqnarray}
\mathbf{b}_D(\theta)&=&[e^{-j\frac{4\pi f_cd\sin\theta}{c}(-\frac{N-1}{2})}, e^{-j\frac{4\pi f_cd\sin\theta}{c}(1-\frac{N-1}{2})},\nonumber\\
&& \dots, e^{-j\frac{4\pi f_cd\sin\theta}{c}(\frac{N-1}{2})}]^T,
\end{eqnarray} 
and
\begin{eqnarray}
\mathbf{b}_R(r)&=&e^{-j\frac{4\pi}{c}f_cr}[e^{-j\frac{4\pi \Delta{f}}{c}m_0r}, e^{-j\frac{4\pi \Delta{f}}{c}m_1r},\nonumber\\
&& \dots, e^{-j\frac{4\pi \Delta{f}}{c}m_{N-1}r}]^T.
\end{eqnarray}
For the multi-target, multi-snapshot, and noisy scenario, suppose the direction and range of the $i$th target is $\{\theta_i, r_i\}$ ($i=1,2,\dots,P$, $P$ is the target number), and the complex reflection amplitude of the $i$th targets at the $l$th snapshot is $\alpha_i(l)$ ($l=0,1\dots, L-1$, $L$ is the snapshot number.). Then the baseband echo vector of the $l$th snapshot is the superimposition of echoes from all targets:
\begin{equation}
\label{Eq:SignalModelMultiSnapshot}
\mathbf{r}(l)=\sum_{i=1}^{P}a_i(l)\cdot\mathbf{b}_D(\theta_i)\odot\mathbf{b}_R(r_i)+\mathbf{v}(l),
\end{equation}
where $\mathbf{v}(l)$ is the $N\times 1$ additive receiver noise vector. In (\ref{Eq:SignalModelMultiSnapshot}), $a_i(l)$ can vary from snapshot to snapshot, because of the target's fluctuation \cite{merrill2001introduction}.

\section{The Beampattern of RFDA}
\label{Sec:BeamPattern}
The system and signal model of RFDA were formulated in the last section. In this section, we will discuss its beampattern, which is an important characteristic of an array antenna  \cite{merrill2001introduction}. In array processing \cite{van2004detection}, the beampattern is the system response of an array beamformed in one direction to a unit amplitude target located in another direction. According to (\ref{Eq:BasebandScalar}), the received echo depends on both the direction and the range simultaneously. So the corresponding beampatterns of RFDAs are functions of both direction and range. The following discussion will show that, with signal processing, RFDA can indicate both the target's direction and range without coupling.

\subsection{Definition and Basic Results}
\label{subsec:BeamPattern:Expression}
\begin{figure*}[!t]
\centering
\includegraphics[width=0.7\textwidth]{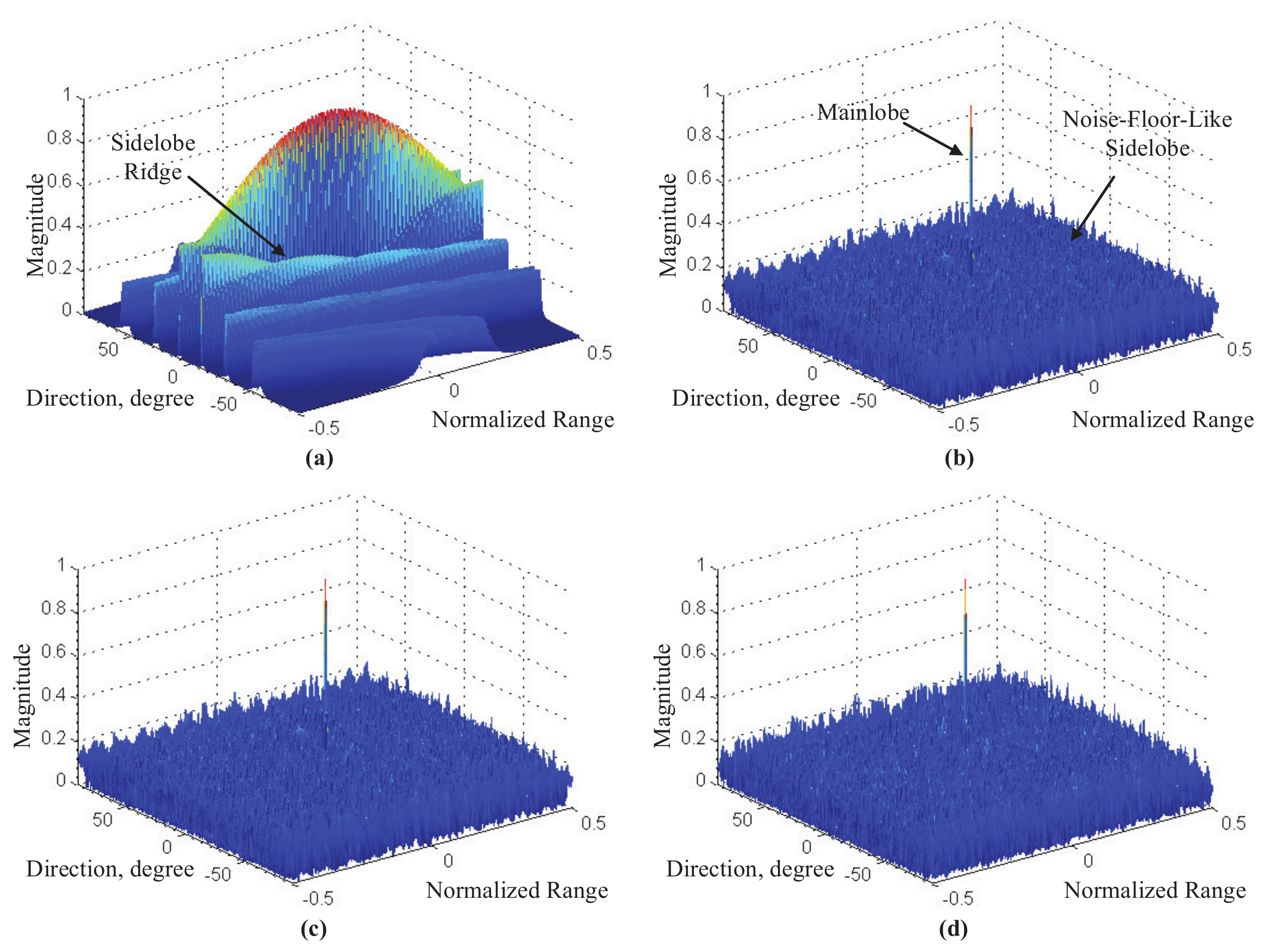}
\caption{Examples of beampatterns of different frequency diverse arrays. (a) LFDA; (b) RFDA, discrete uniform distributed; (c) RFDA, continuous uniform distributed; (d) RFDA, Gaussian distributed.}
\label{Fig:BeamFDA}
\end{figure*}

If the target is located at $\{\theta_1, r_1\}$, and the RFDA is beamformed to another location, named $\{\theta_2, r_2\}$, the array response is
\begin{equation}
\label{Eq:BeamPatternDefinition}
\beta(\{\theta_1,r_1\},\{\theta_2,r_2\})=\frac{<\mathbf{b}(\theta_1,r_1), \mathbf{b}(\theta_2,r_2)>}{\|\mathbf{b}(\theta_1,r_1)\|_2\|\mathbf{b}(\theta_2,r_2)\|_2}.
\end{equation}

Substituting (\ref{Eq:BasebandScalar}) and (\ref{Eq:SteeringVector}) into (\ref{Eq:BeamPatternDefinition}), the beampattern of a RFDA is
\begin{eqnarray}
\label{Eq:BeamPatternDifference}
&&\beta(\{\theta_1,r_1\},\{\theta_2,r_2\})\nonumber\\
&&=\frac{1}{N}[\mathbf{b}_D(\theta_1)\odot\mathbf{b}_R(r_1)]^H[\mathbf{b}_D(\theta_2)\odot\mathbf{b}_R(r_2)]\nonumber\\
&&=\frac{1}{N}[\mathbf{b}_D^*(\theta_1)\odot\mathbf{b}_D(\theta_2)]^T[\mathbf{b}_R^*(r_1)\odot\mathbf{b}_R(r_2)]\nonumber\\
&&=\frac{1}{N}e^{j2\pi\frac{p}{\delta}}\sum_{n=0}^{N-1}e^{j2\pi(n-\frac{N-1}{2})q}\cdot e^{j2\pi m_{n}p}\nonumber\\
&&=\beta(q,p),
\end{eqnarray}
where $[\cdot]^*$ indicates element-wise conjugation. The variables $q$, $p$ and $\delta$ are defined as $q=2(\sin\theta_1-\sin\theta_2)f_cd/c$, $p=2(r_1-r_2)\Delta{f}/c$, and $\delta=\Delta{f}/{f_c}$. According to (\ref{Eq:BeamPatternDifference}), the RFDA beampattern depends only on the difference of direction sine, $\sin\theta_1-\sin\theta_2$, and the difference of range , $r_1-r_2$. But it is independent of the absolute value of the target's direction or range.

Meanwhile, for traditional  LFDAs \cite{Antonik2006Frequency}\cite{sammartino2013frequency}, the transmitted frequencies are linearly shifted along the element. Hence the signal models of LFDA can be achieved by alternating $m_n$ with $n-\frac{N-1}{2}$ in (\ref{Eq:BeamPatternDifference}),
\begin{equation}
\label{Eq:BeamPatternSimpleLinear}
\beta(q,p)=\frac{1}{N}e^{j2\pi\frac{p}{\delta}}\sum_{n=0}^{N-1}e^{j2\pi(p+q)\left(n-\frac{N-1}{2}\right)}.
\end{equation}

Fig. \ref{Fig:BeamFDA} shows the beampatterns of different kinds of FDA antennas. Subfigure (a) is the beampattern of an LFDA. As shown, this beampattern has high sidelobe ridges, which implies that the direction and range are coupled, and the indication of target location is ambiguous. The other three subfigures are the beampatterns of RFDA antennas. Subfigure (b) is the beampattern when the carrier frequency of each element is discrete uniform distributed, while the beampatterns of (c) and (d) are continuous uniform and Gaussian, respectively. However, regardless of the frequency distribution, the RFDAs' beampatterns are thumbtack-like, and all the peaks are located where  $p=0$ and $q=0$, which implies that the direction and range have successfully been decoupled and the target location can be uniquely and correctly indicated.

In (\ref{Eq:BeamPatternDifference}), it is shown that the beampattern depends on the distribution of $m_n$ and can be regarded as a stochastic process w.r.t. $q$ and $p$. Hence an analysis about its stochastic features will be very helpful for a more comprehensive study of RFDA's characteristics. In the following  subsections, we will explicitly derive expressions of the mean, variance, and asymptotic distribution of the beampattern.

\subsection{Mean}
\label{Subsec:MeanBeamPattern}

The mean of the beampattern is derived in this subsection. Because the transmitted frequencies of each element are \textit{i.i.d.}, the mean of the beampattern can be calculated as
\begin{eqnarray}
\label{Eq:MeanBeamDef}
\bar{\beta}(q,p)&\triangleq& \mathbb{E}_{\mathbf{m}}\left\{\beta(q,p)\right\}\nonumber\\
&=&\int_{m_n\in\mathcal{M}} \beta(q,p) g(m_n)dm_n\nonumber\\
&=&\frac{1}{N}e^{j2\pi\frac{p}{\delta}}\sum_{n=0}^{N-1}e^{j2\pi\left(n-\frac{N-1}{2}\right)q}\nonumber\\
&&\cdot\int g(m_n)e^{j2\pi m_np}dm_n\nonumber\\
&=&\frac{1}{N}e^{j2\pi\frac{p}{\delta}}S_a^N(q)\cdot\Phi(p),
\end{eqnarray}
where $\mathcal{M}$ is the sample space of $m_n$, $S_a^N(x)=(\sin N\pi x)/(N\sin \pi x)$, and $\Phi(x)$ is the moment-generating function of $m_n$,
\begin{equation}
\label{Eq:CharactorFunction}
\Phi(x)=\int_{m_n\in\mathcal{M}} g(m_n)e^{j2\pi m_nx}dm_n.
\end{equation}

For the three carrier frequency distribution instances, when $m_n$ is Gaussian distributed, $\Phi(x)=e^{-2\pi^2\sigma^2x^2}$ ($\sigma^2$ is the variance), and then the mean of the beampattern is
\begin{equation}
\label{Eq:MeanBeamGaussian}
\bar{\beta}_G(q,p)=\frac{1}{N}e^{j2\pi\frac{p}{\delta}}S_a^N(q)\cdot e^{-2\pi^2\sigma^2p^2}.
\end{equation}
Besides, for continuous and discrete uniform distributions, $\bar{\beta}(q,p)$ can be achieved through substituting $g(m_n)={1}/{M}$, $m_n\in[-{M}/{2},{M}/{2}]$, and $g(m_n)={1}/{M}$, $m\in\{-({M-1})/{2},-(M-1)/{2}+1,\dots, (M-1)/{2}]$ into (\ref{Eq:MeanBeamDef}-\ref{Eq:CharactorFunction}), respectively as
\begin{equation}
\label{Eq:MeanBeamContinUni}
\bar{\beta}_C(q,p)=\frac{1}{N}e^{j2\pi\frac{p}{\delta}}S_a^N(q)\cdot \frac{\sin(M\pi p)}{M\pi p},
\end{equation}
and
\begin{equation}
\label{Eq:MeanBeamDiscreteUni}
\bar{\beta}_D(q,p)=\frac{1}{MN}e^{j2\pi\frac{p}{\delta}}S_a^N(q)\cdot S_a^M(p).
\end{equation}

According to (\ref{Eq:MeanBeamDef}), in the mean of the beampattern, the direction variable $q$ and range variable $p$ are decoupled. Moreover, it is known that the direction beampattern of a traditional $N$-element linear array is
\begin{equation}
\beta(q)=\frac{\sin N\pi q}{N\sin \pi q}=S_a^N(q),\nonumber
\end{equation}
and the range response  (the ``range beampattern") of a random frequency radar is 
\begin{equation}
\beta(p)=\int_{m_n\in\mathcal{M}} g(m_n)e^{j2\pi m_np}dm_n=\Phi(p).\nonumber
\end{equation}
Hence the mean of an RFDA's beampattern can be regarded as the Cartesian product of the direction beampattern and range beampattern, which also implies that the RFDA has the ability to resolve the target direction and range simultaneously. Moreover, the direction and range resolution of RFDA can be respectively evaluated by $S_a^N(q)$ and $\Phi(p)$.

\begin{figure}[!h]
\centering
\includegraphics[width=.5\textwidth]{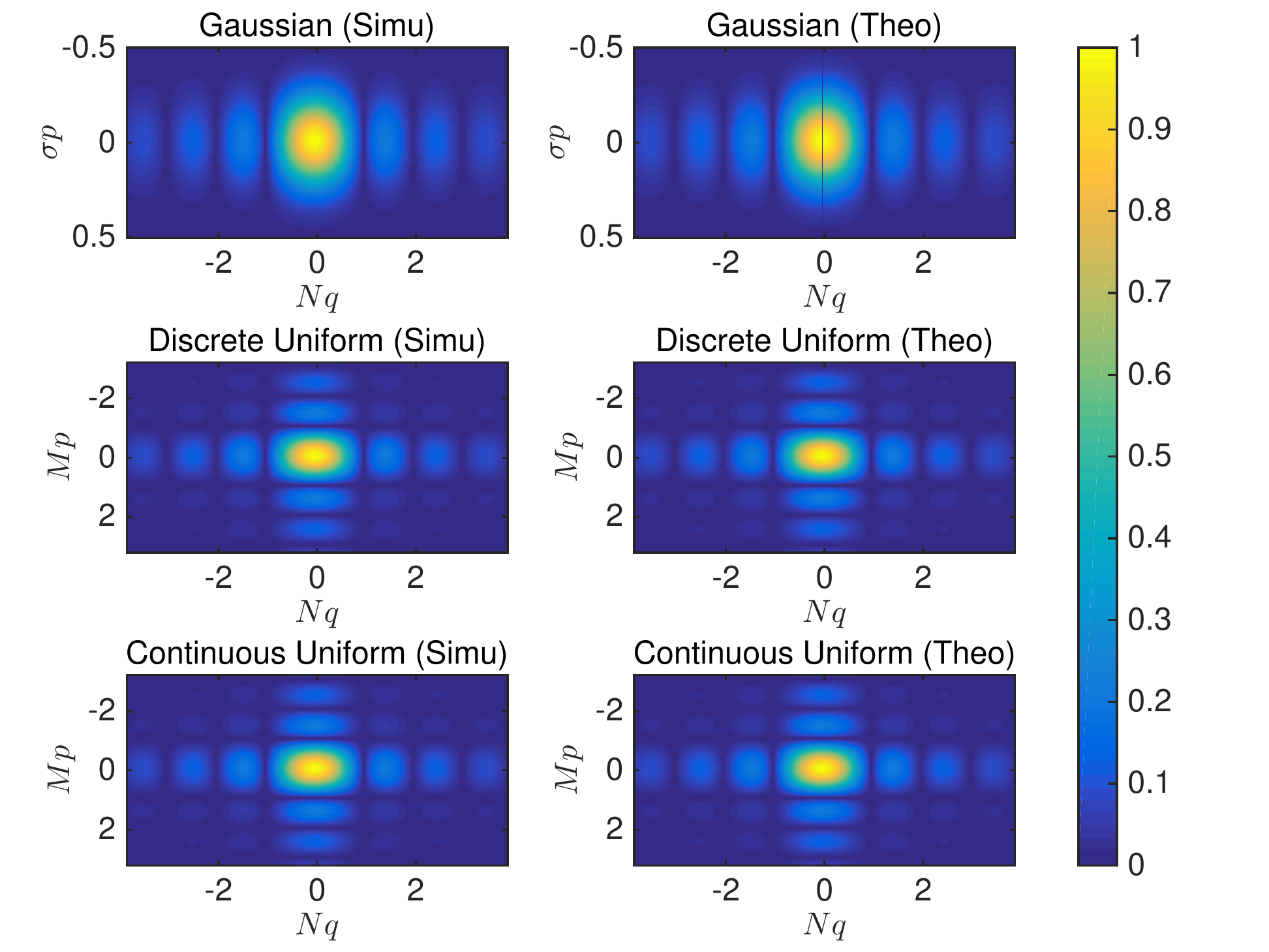}
\caption{The mean of beampatterns (in amplitude). The figures in the left column are the averaged results of 10,000 Monte Carlo trials. The right column shows the theoretical values which are plotted following the expressions in (\ref{Eq:MeanBeamGaussian}-\ref{Eq:MeanBeamDiscreteUni}). The upper, middle, and lower rows are for the Gaussian, discrete, and continuous uniformly distributed carrier frequencies, respectively}
\label{Fig:BeamMean}
\end{figure}

Fig. \ref{Fig:BeamMean} gives an example of the beampattern mean of RFDAs with different kinds of distributions. The averaged beampattern of 10,000 Monte Carlo trials (in each trial, the carrier frequencies of all the array elements are independent samples from the corresponding distribution.) are listed in the left column, and the theoretical results are shown in the right column. The comparison shows that,  for all the Gaussian (upper row and (\ref{Eq:MeanBeamGaussian})), discrete uniform (lower row and (\ref{Eq:MeanBeamDiscreteUni})), and continuous uniform (middle row and (\ref{Eq:MeanBeamContinUni})) distributions, the simulation results match well with the theoretical expressions. 


\subsection{Variance}
\label{Subsec:VarBeamPattern}
Fig. \ref{Fig:BeamFDA} shows that the beampatterns of RFDA antennas have a noise-like sidelobe base. In single target scenarios, the sidelobe base is immaterial for target detection and direction-range indication. But in multi-target scenarios, the sidelobe base of dominant targets may conceal weak targets. Hence it is necessary to analysis the peak-sidelobe-base-ratio (PSBR).

As introduced in Subsection \ref{subsec:BeamPattern:Expression}, the beampattern can be regarded as a random process. Hence the PSBR corresponding to each $\{q, p\}$ pair can be measured by the inverse proportion of the beampattern's variance, whose explicit expression can be provided by the following derivation.
\begin{eqnarray}
\label{Eq:VarBeamPatternDef}
\sigma^2_{\beta}(q,p)&\triangleq&\mathbb{V}_{\mathbf{m}}\left\{\beta(q,p)\right\}\nonumber\\
&=&\mathbb{E}_{\mathbf{m}}\left\{\beta^*(q,p)\beta(q,p)\right\}-|\bar{\beta}(q,p)|^2\nonumber\\
&=&\frac{1}{N}-\frac{1}{N}|\Phi(p)|^2.
\end{eqnarray}

Equation (\ref{Eq:VarBeamPatternDef}) shows that the PSBR of an RFDA depends on the element number, $N$, and the moment-generating function, $\Phi(p)$. Furthermore, this equation also implies that
\begin {enumerate}
\item
The variance $\sigma^2_{\beta}(q,p)$ is a function of $p$, which means that the PSBR is determined only by the range. Different directions with the same range have the same PSBR;
\item
When $|\Phi(p)|$ is sufficiently small, the PSBR approaches ${N}$, which means the larger the element number, the larger the PSBR. It will also be easier to unmask weak targets from sidelobes of dominant targets.
\item
The variance $\sigma^2_{\beta}(q,p)$ is equal to zero when $p=0$, which means the beampattern is deterministic when the range difference is zero, and
\begin{equation}
\label{Eq:DeteministicBeamPattern}
\beta(q,0)=\bar{\beta}(q,0)=\frac{1}{N}S_a^N(q).
\end{equation}
\end{enumerate}

\begin{figure}[!h]
\centering
\includegraphics[width=.5\textwidth]{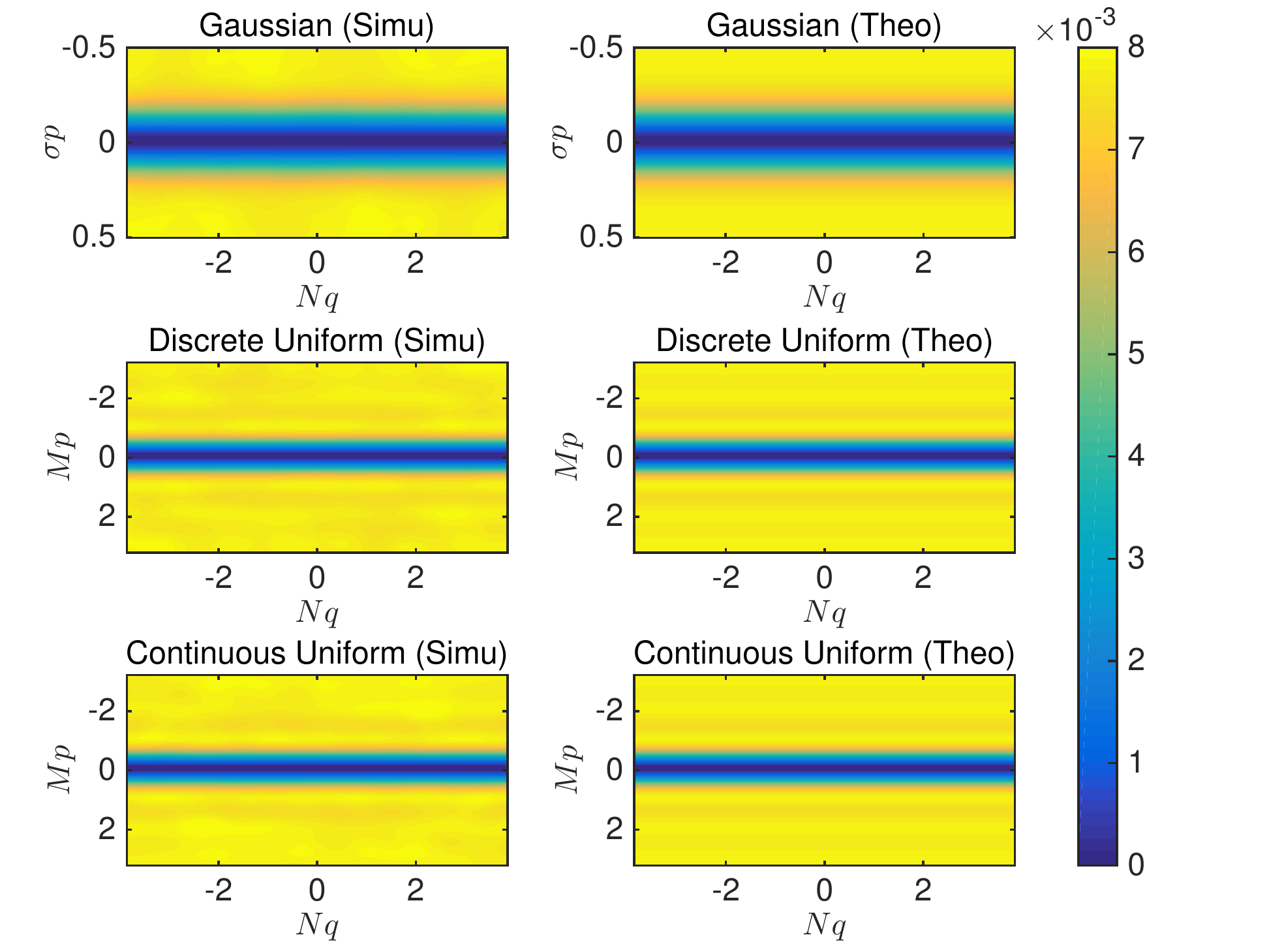}
\caption{The variances of beampatterns. All the subfigures in the left column are counted from 10,000 Monte Carlo trials. And the right column shows the theoretical results provided in (\ref{Eq:VarBeamPatternDef}). The upper, middle and lower rows are the results for Gaussian, discrete, and continuous uniformly distributed carrier frequencies, respectively.
}
\label{Fig:BeamVar}
\end{figure}

The variances of beampatterns are illustrated in Fig. \ref{Fig:BeamVar}. All the simulated variances, which are counted from the results of 10,000 Monte Carlo trials, are displayed in the left column, and the theoretical values given by (\ref{Eq:VarBeamPatternDef}) are on the right. This illustration shows that, for the variances of beampatterns, the simulation results also match the theoretical expressions well.

\subsection{Asymptotic Distribution}
\label{Subsec:AsymBeamPattern:}
In subsections \ref{Subsec:MeanBeamPattern} and \ref{Subsec:VarBeamPattern}, the mean and variance of the beampattern are given to evaluate the resolution and PSBR of RFDA. An exact distribution of the beampattern is preferred if more properties are required. However, it is quite difficult to give an explicit form of the beampattern's distribution. Alternatively, we will give an asymptotic distribution of the beampattern, which is an acceptable approximation when the number of array elements is sufficiently large. Some of the derivations are inspired by \cite{lo1964mathematical}, and we expand the approach to a 2D case for RFDA.

According to (\ref{Eq:BeamPatternDifference}), $\beta(q,p)$ can be interpreted as a sum of $N$ independent random variables $y_n(q,p)$, where
\begin{equation}
\label{Eq:ExpressionofYn}
y_n(q,p)=\frac{1}{N}e^{j2\pi\frac{p}{\delta}}e^{j2\pi(n-\frac{N-1}{2})q}e^{j2\pi m_np},
\end{equation}
and
\begin{equation}
\label{Eq:SumofRandomV}
\beta(q,p)=\sum_{n=0}^{N-1}y_n(q,p).
\end{equation}
Then the mean and variance of $y_n(q,p)$ are
\begin{eqnarray}
\label{Eq:MeanYn}
\mathbb{E}_{m_n}\left\{y_n(q,p)\right\}&=&\int_{m_n\in\mathcal{M}}y_n(q,p) g(m_n)dm_n\nonumber\\
&=&\frac{1}{N}e^{j2\pi(n-\frac{N-1}{2})q}e^{j2\pi\frac{p}{\delta}}\Phi(p),
\end{eqnarray}
\begin{eqnarray}
\label{Eq:VarYn}
\mathbb{V}_{m_n}\{y_n(q,p)\}&=&\mathbb{E}_{m_n}\{|y_n(q,p)|^2\}-|\mathbb{E}_{m_n}\{y_n(q,p)\}|^2\nonumber\\
&=&\frac{1}{N^2}-\frac{1}{N^2}|\Phi(p)|^2.
\end{eqnarray}

It can be validated that $y_n(q,p)$ can satisfy the sufficient condition of the Lyapunov central limit theorem \cite{kolmogorov1954limit}. Hence the normalized sum of $y_n(q,p)$, given by
\begin{equation}
\label{Eq:LCT}
\sum_{n=0}^{N-1}\frac{y_n(q,p)-\mathbb{E}_{m_n}\{y_n(q,p)\}}{\sqrt{\mathbb{V}_{m_n}\{y_n(q,p)\}}},\nonumber
\end{equation}
approaches the standard normal distribution when the element number $N$ is sufficiently large.

So according to the Lyapunov central limit theorem, the asymptotic distribution of $\beta(q,p)$ can be regarded as complex Gaussian. Denote $\beta_1(q,p)$, $\beta_2(q,p)$ as the real and imaginary parts of the $\beta(q,p)$, and we have the following theorem.

\begin{theorem}
\label{Th:AsyDistribution}
Random vector ${\bm{\beta}}(q,p)\triangleq[\beta_1(q,p)$, $\beta_2(q,p)]^T$ is asymptotically joint Gaussian distributed. The mean and covariance matrix of $\bm{\beta}(q,p)$ are
\begin{equation}
\label{Eq:MeanofAsy}
\mathbb{E}_{\mathbf{m}}\left\{{\bm{\beta}}(q,p)\right\}=\frac{1}{N}S_a^N(q)\Phi(p)\left[\begin{array}{c}\cos{\alpha}\\\sin{\alpha}\end{array}\right],
\end{equation}
and
\begin{eqnarray}
\label{Eq:CovofAsy}
&&\mathbf{M}_{{\bm{\beta}}}(q,p)\nonumber\\
&&\triangleq \mathbb{E}_{\mathbf{m}}\Big\{[{\bm{\beta}}(q,p)-E\{{\bm{\beta}}(q,p)\}][{\bm{\beta}}(q,p)-E\{{\bm{\beta}}(q,p)\}]^T\Big\}\nonumber\\
&&=\left[\begin{array}{cc}\sigma_r^2\cos^2\alpha+\sigma_i^2\sin^2\alpha&(\sigma_r^2-\sigma_i^2)\sin\alpha\cos\alpha\\(\sigma_r^2-\sigma_i^2)\sin\alpha\cos\alpha&\sigma_r^2\sin^2\alpha+\sigma_i^2\cos^2\alpha\end{array}\right],\nonumber\\
\end{eqnarray}
where $\alpha={2\pi p/\delta}$, and
\begin{equation}
\sigma_r^2=\frac{1}{2N}\left[1-\Phi^2(p)-\frac{S_a^N(2q)}{N}\left(\Phi^2(p)-\Phi(2p)\right)\right],
\end{equation}
\begin{equation}
\sigma_i^2=\frac{1}{2N}\left[1-\Phi^2(p)+\frac{S_a^N(2q)}{N}\left(\Phi^2(p)-\Phi(2p)\right)\right].
\end{equation}
\end{theorem}

\textbf{Proof:} The proof can be found in Appendix. \hfill{$\square$}

With the asymptotic distribution of the beampattern, we can derive Proposition\ref{Pr:Sidelobe} for the sidelobe level of the RFDA. 
\begin{corollary}
\label{Pr:Sidelobe}
For a direction difference $q={1}/{2N}$, the complementary cumulative distribution function of the sidelobe magnitude at $\{q,p\}$ satisfies
\begin{equation}
\mbox{Pr}\{|\beta(q,p)|>r\}=Q_1\left(\frac{a}{\tau}, \frac{r}{\tau}\right),
\end{equation}
where $Q_1(x,y)$ is the first-order Marcum Q-function, 
\begin{equation}
Q_1(x,y)=\int_{b}^{\infty}te^{-\frac{t^2+x^2}{2}}I_0(xt)dt,\nonumber\\
\end{equation}
and $I_0(\cdot)$ is the modified Bessel function of the first kind of zero order, and
\begin{eqnarray}
a&=&\frac{1}{N}\left|S_a^N(q)\Phi(p)\right|,\nonumber\\
\tau&=&\sqrt{\frac{1-\Phi^2(p)}{2N}}.\nonumber
\end{eqnarray}
\end{corollary}

\textbf{Proof:} The corollary can be achieved directly by combining Theorem \ref{Th:AsyDistribution} and equation (2.18) of \cite{simon2007probability}. \hfill{$\square$}

\section{Signal Processing Methods}
\label{sec:SignalProcessing}
The beampattern provided in last section shows that, in RFDA, the target direction and range can be decoupled. In this section, we will introduce two signal processing methods for target indication and direction/range estimation.

\subsection{Matched Filtering}
\label{Subsec:BF}
Matched filtering is an effective way to obtain the highest filtered SNR and the intrinsic resolution of waveforms \cite{rihaczek1969principles}. According to the equivalence between matched filtering results and the beampattern \cite{van2004detection}, the uncoupled direction-range indication and direction/range resolution can be achieved by matching filtering, especially in single target scenarios.

In this paper, we omit the original matching-filtering algorithm because it is straightforward. However, for the discrete uniform distribution of carrier frequencies, if the relative  bandwidth, $M\Delta{f}/f_c$, is small, another computation effective alternative, named zero-padding and 2D fast Fourier transformation (zero-padding-2DFFT), can be employed to implement of matched filtering. The main steps of this method are
\begin {enumerate}
\item
\textit{Step 1}: Formulate an all-zero $M\times N$ data matrix.
\item
\textit{Step 2}: For all the $n=0,1,\dots, N-1$, fill the $\big(m_n+(M-1)/{2}\big)$th row, $n$th column entry with the baseband echo sample $b_n$, and leave all the other entries zero.
\item
\textit{Step 3}: Apply two 2DFFT on the filled data matrix, and the output can approximate the matched filtering result. A finer result will be achieved via FFTs with more points. 
\end {enumerate}

\begin{figure}[!t]
\centering
\includegraphics[width=.5\textwidth]{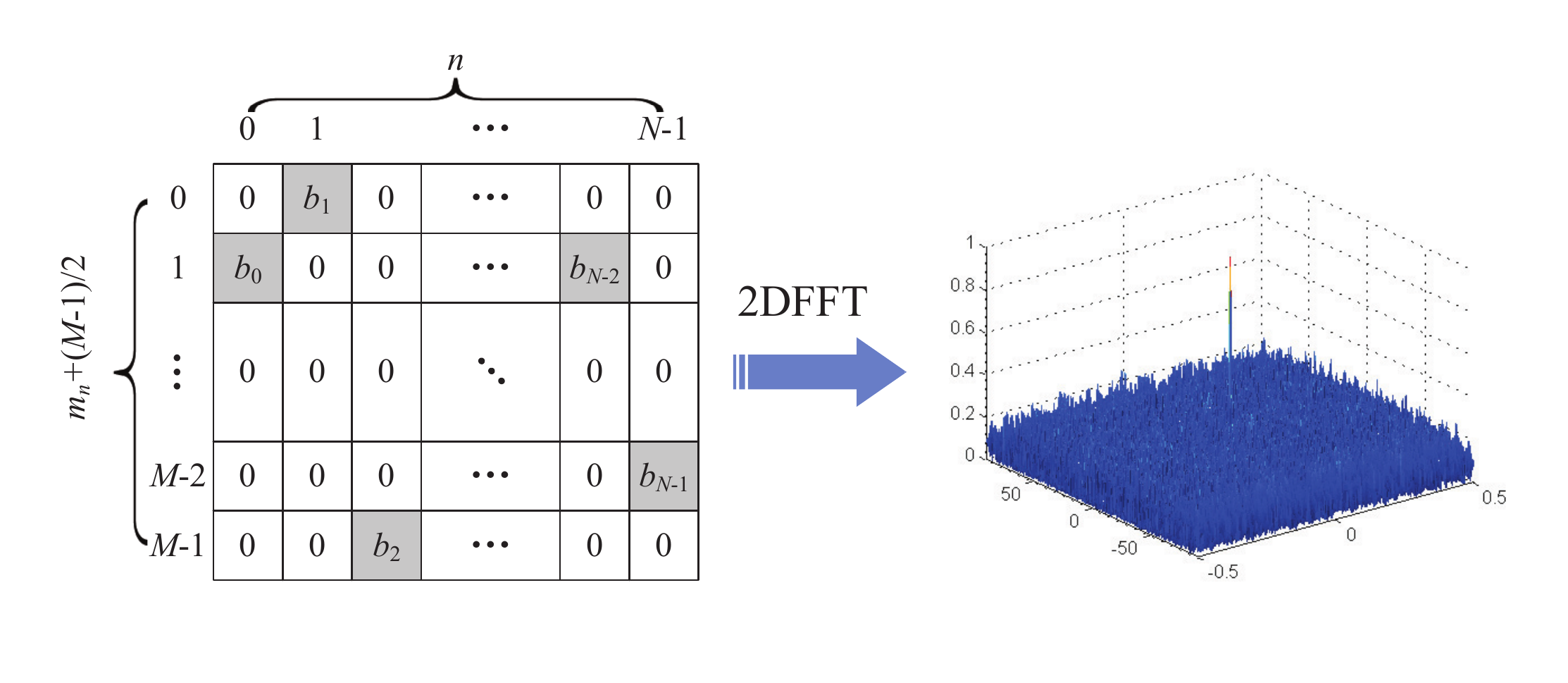}
\caption{A brief illustration of the zero-padding-2DFFT method, where $m_0=-(M-3)/2, m_1=-(M-1)/2,\dots, m_{N-1}={(M-3)}/{2}$.}
\label{Fig:2DFFT}
\end{figure}

The zero-padding-2DFFT method and its results are illustrated in Fig. \ref{Fig:2DFFT}. With the effectiveness of FFT, the new proposed algorithm can reduce the computation load from $O(MN^2)$ (due to the inner products between the received signal vector $\mathbf{r}$ and the array steering vectors of all the possible ranges and directions.) to $O(NM\cdot\log M+MN\cdot\log N)$. 

\subsection {Compressive Sensing for Direction and Range Indication}
\label{Subsec:CS}
As mentioned, in multi-target scenarios, the strong targets' sidelobe bases may mask weak targets. In this subsection, we adopt the sparse recovery algorithms for compressive sensing to solve this problem.

First, the unambiguous extent of direction and unambiguous extent of range are uniformly divided into $P$ and $Q$ units, respectively (For the ambiguous direction and range problem, the reader can refer to \cite{sammartino2013frequency} and \cite{xu2015joint} for more information.). There will be $PQ$ direction-range pairs, $\{\theta_i, r_i\}_{i=1}^{PQ}$, in the unambiguous range-direction extent.

Define a $PQ\times 1$ vector $\mathbf{x}(l)$, whose $i$th entry is the reflection magnitude of a target located at $\{\theta_i, r_i\}$, at the $l$th snapshot. The observing matrix $\mathbf{\Phi}$ is $N\times PQ$, and its $i$th column is determined by (\ref{Eq:SteeringVector}). Then in the single snapshot case (named the single measurement vector (SMV) in the compressive sensing realm), the noiseless echo can be rewritten in a matrix form:
\begin{equation}
\label{Eq:SMV_Noiseless}
\mathbf{r}(l)=\mathbf{\Phi}\mathbf{x}(l).
\end{equation}
And if the observations are noisy, an additive noise vector $\mathbf{v}$ needs to be added on the right side of above equation, as
\begin{equation}
\label{Eq:SMV_Noisy}
\mathbf{r}(l)=\mathbf{\Phi}\mathbf{x}(l)+\mathbf{v}.
\end{equation}

If there are $L$ ($L>1$) snapshots, the noisy echoes can be expressed for the multiple measure vector (MMV) scenario as
\begin{equation}
\label{Eq:MMV}
\mathbf{R}=\mathbf{\Phi}\mathbf{X}+\mathbf{V},
\end{equation}
where $\mathbf{R}=[\mathbf{r}(1), \mathbf{r}(1),\dots, \mathbf{r}(L)]$, $\mathbf{X}=[\mathbf{x}(1), \mathbf{x}(1),\dots, \mathbf{x}(L)]$, and $\mathbf{V}$ is the receiver noise matrix.

For the noiseless case, the compressive sensing can be accomplished by basis pursuit (BP) \cite{chen1998atomic} for sparse recovery of the targets:
\begin{equation}
\label{Eq:BP}
\hat{\mathbf{x}}(l)=\argmin_{\mathbf{x}(l)}\|\mathbf{x}(l)\|_1, \mbox{ s.t. }\mathbf{y}=\mathbf{\Phi}\mathbf{x}.
\end{equation}
And for the noisy case, the BP changes to quadratic constrained basis pursuit (QCBP) \cite{malioutov2005sparse}:
\begin{equation}
\label{Eq:QCBP}
\hat{\mathbf{x}}(l)=\argmin_{\mathbf{x}(l)}\|\mathbf{x}(l)\|_1, \mbox{ s.t. }\|\mathbf{y}-\mathbf{\Phi}\mathbf{x}(l)\|_2<\sigma_l.
\end{equation}

Greedy methods, such as subspace pursuit (SP) \cite{dai2009subspace}, are attractive alternatives for reducing the computation load while maintaining the recovery quality. 
For the SMV scenario, the sparsest estimate of $\mathbf{x}(l)$ can be achieved by
\begin{equation}
\label{Eq:GreedyforSMV}
\min \|\mathbf{x}(l)\|_0, \mbox{ subject to }\|\mathbf{r}(l)-\mathbf{\Phi}\mathbf{x}(l)\|_2\leq\sigma_l.
\end{equation}

For the MMV scenario, we choose the $\|\cdot\|_{2,0}$ norm to maintain the consistency of the targets' locations in all the snapshots. The estimate of $\mathbf{X}$ is 
\begin{equation}
\label{Eq:GreedyforMMV}
\min \|\mathbf{X}\|_{2,0}, \mbox{ subject to }\|\mathbf{R}-\mathbf{\Phi}\mathbf{X}\|_F\leq\sigma_L,
\end{equation}
where $\|\cdot\|_F$ is the Frobenius norm. In (\ref{Eq:GreedyforSMV}) and (\ref{Eq:GreedyforMMV}), $\sigma_l$ and $\sigma_L$ are the error tolerances determined by the noise power.

In compressive sensing, correct recovery can be guaranteed with high probability if the observing matrix $\mathbf{\Phi}$ has a small mutual coherence \cite{ben2010coherence}. Mutual coherence is defined as the maximal normalized inner product between different columns of the observing matrix
\begin{equation}
\label{Eq:CoherenceDef}
\mu\triangleq\max_{i\neq h}\frac{\big|[\mathbf{\Phi}]_i^H[\mathbf{\Phi}]_h\big|}{\big\|[\mathbf{\Phi}]_i\big\|_2\big\|[\mathbf{\Phi}]_h\big\|_2}.
\end{equation}
As with (\ref{Eq:BeamPatternDefinition}), the mutual coherence is equal to the highest sidelobe in the beampattern. The result in Fig. \ref{Fig:BeamFDA} shows that, in comparison with the LFDA, the highest sidelobe level of the RFDA is fairly low. This property means that the RFDA, which can be regarded as a random sparse sampling in the spatial-frequency domain, is suitable for compressive sensing. An investigation of the RFDA's mutual coherence will be provided in Subsection \ref{sub:Coherence}. 

In this work, the SP, FOCUSS \cite{gorodnitsky1997sparse}, and their MMV extensions, the Generalized Subspace Pursuit (GSP) \cite{Feng2013Generalized} and the M-FOCUSS  \cite{cotter2005sparse}, are adopted to recover the targets in both SMV and MMV scenarios. The demonstration and performance comparison of different sparse recovery algorithms will be given in Section \ref{sec:Results}.

\section{Performance Analysis}
\label{Sec:PerformanceAnalysis}
In previous sections, it was shown that the direction and range are decoupled in the beampattern of an RFDA, and the directions and ranges of targets can be estimated via matched filtering or compressive sensing. In this section, we will provide performance bounds of the RFDA, by deriving the CRB of the estimation error and the mutual coherence of the observing matrix. The CRB can be used to evaluate the location accuracy, and the mutual coherence is a helpful metric to determine the sparse recovery performance.

\subsection{Cram\'{e}r-Rao Bound}
\label{subsec:CRB}
Because the mean square errors (MSEs) of the un-biased estimation are lower bounded by the CRB \cite{kay1998fundamentals}, we will provide the CRB of the direction/range estimation, which can be adopted as a performance guarantee of the RFDA.

Since the target number is $P$, the directions and ranges of all the targets can be combined into a $2P\times 1$ parameter vector, $\bm{\xi}=[\theta_1, r_1, \theta_2, r_2 \dots, \theta_P, r_P]^T$.  According to the signal model in (\ref{Eq:SignalModelMultiSnapshot}), and assuming the receiver noise is complex additive-white-Gaussian-noise (AWGN), the logarithmic likelihood function of $\bm{\xi}$ is
\begin{equation}
\label{Eq:LogLikelihood}
\mathcal{L}\left(\mathbf{r}(l);\bm{\xi}\right)=C-\frac{1}{\sigma_n^2}\sum_{l=0}^{L-1}\Big|\mathbf{\mathbf{r}}(l)-\sum_{i=1}^{P}\alpha_i(l)\mathbf{b}_D(\theta_i)\odot\mathbf{b}_R(r_i)\Big|^2,
\end{equation}
where $C$ is a constant and $\sigma_n^2$ is the receiver  noise power.

Then the MSE matrix of $\hat{\bm{\xi}}$ ($\hat{\bm{\xi}}$ is an unbiased estimation of $\bm{\xi}$) satisfies the information inequality \cite{kay1998fundamentals}
\begin{equation}
\label{Eq:FIM}
E\left\{(\hat{\bm{\xi}}-\bm{\xi})(\hat{\bm{\xi}}-\bm{\xi})^{H}\right\}\succeq \mathbf{J}_{\bm{\xi}}^{-1},
\end{equation}
where $\mathbf{J}_{\bm{\xi}}$ is the Fisher information matrix (FIM) of $\bm{\xi}$.

With the results in \cite{stoica1990performance}, the FIM can be calculated by
\begin{equation}
\label{Eq:FIMCal}
\mathbf{J}_{\bm{\xi}}=\frac{2L}{\sigma_n^2}\Re\left\{\mathbf{C}\odot\left(\mathbf{S}^T\otimes\left[\begin{array}{cc}1&1\\1&1\end{array}\right]\right)\right\},
\end{equation}
where $\mathbf{S}$ is a $P\times P$ matrix, whose $i$, $j$th entry is
\begin{equation}
[\mathbf{S}]_{ij}=\frac{1}{L}\sum_{l=1}^{L}\alpha_i(l)\alpha_j^{*}(l),
\end{equation}
and
\begin{equation}
\label{Eq:DefC}
\mathbf{C}=\mathbf{D}^H\mathbf{P}_{\mathbf{A}}^{\perp}\mathbf{D}=(\mathbf{P}_{\mathbf{A}}^{\perp}\mathbf{D})^H\mathbf{P}_{\mathbf{A}}^{\perp}\mathbf{D}.
\end{equation}
In (\ref{Eq:DefC}),
\begin{eqnarray}
\mathbf{D}&=&\Big[\frac{\partial \mathbf{b}(\theta_1, r_1)}{\partial \theta_1}, \frac{\partial \mathbf{b}(\theta_1,r_1)}{\partial r_1}, 
\frac{\partial \mathbf{b}(\theta_2,r_2)}{\partial \theta_2}, \frac{\partial \mathbf{b}(\theta_2, r_2)}{\partial r_2}, \nonumber\\
&&\dots, \frac{\partial \mathbf{b}(\theta_P,r_P)}{\partial \theta_P}, \frac{\partial \mathbf{b}(\theta_P,r_P)}{\partial r_P}\Big],
\end{eqnarray}
and $\mathbf{A}$ is an $N \times P$ matrix, given by
\begin{equation}
\mathbf{A}=[\mathbf{b}(\theta_1, r_1), \mathbf{b}(\theta_2, r_2), \dots, \mathbf{b}(\theta_P,r_P)],
\end{equation}
where $\mathbf{P_{\mathbf{A}}^{\perp}}$ is the projection matrix onto the orthogonal complement of $\mathbf{A}$'s column space:
\begin{equation}
\mathbf{P_{\mathbf{A}}^{\perp}}=\mathbf{I}-\mathbf{A}(\mathbf{A}^H\mathbf{A})^{-1}\mathbf{A}^H,
\end{equation}
and $\mathbf{I}$ is the identity matrix.

According to the above definitions, $\mathbf{J}_{\bm{\xi}}$ is a block matrix with $P\times P$ blocks, and
\begin{eqnarray}
\label{Eq:BlockFIM}
&&\mathbf{J}_{\bm{\xi}}=\frac{2L}{\sigma_n^2}\nonumber\\
&&\cdot\Re\Big\{\left[\begin{array}{cccc}
\left[\mathbf{S}\right]_{11}\mathbf{G}_{11}&\left[\mathbf{S}\right]_{12}\mathbf{G}_{12}&\cdots&\left[\mathbf{S}\right]_{1P}\mathbf{G}_{1P}\\
\left[\mathbf{S}\right]_{21}\mathbf{G}_{21}&\left[\mathbf{S}\right]_{22}\mathbf{G}_{22}&\cdots&\left[\mathbf{S}\right]_{2P}\mathbf{G}_{2P}\\
\vdots&\vdots&\ddots&\vdots\\
\left[\mathbf{S}\right]_{P1}\mathbf{G}_{P1}&\left[\mathbf{S}\right]_{P2}\mathbf{G}_{P2}&\cdots&\left[\mathbf{S}\right]_{PP}\mathbf{G}_{PP}
\end{array}
\right]\Big\}.\nonumber\\
\end{eqnarray}
In (\ref{Eq:BlockFIM}), each $\mathbf{G}_{ij}$ is a $2\times 2$ block, and its entries can be calculated through
\begin{equation}
\begin{array}{ccc}
\left[\mathbf{G}_{ij}\right]_{11}&=&\mathbf{p}^H_{\theta_i}\mathbf{p}_{\theta_j},\\
\left[\mathbf{G}_{ij}\right]_{12}&=&\mathbf{p}^H_{\theta_i}\mathbf{p}_{r_j},\\
\left[\mathbf{G}_{ij}\right]_{21}&=&\mathbf{p}^H_{r_i}\mathbf{p}_{\theta_j},\\
\left[\mathbf{G}_{ij}\right]_{22}&=&\mathbf{p}^H_{r_i}\mathbf{p}_{r_j},
\end{array}
\end{equation}
where
\begin{equation}
\mathbf{p}_{\theta_i}=\mathbf{P_{\mathbf{A}}^{\perp}}\frac{\partial \mathbf{b}(\theta_i, r_i)}{\partial \theta_i}=-j\frac{4\pi f_cd\cos\theta}{c}\mathbf{P_{\mathbf{A}}^{\perp}}[\mathbf{n}\odot\mathbf{b}(\theta_i, r_i)],\nonumber
\end{equation}
and
\begin{equation}
\mathbf{p}_{r_i}=\mathbf{P_{\mathbf{A}}^{\perp}}\frac{\partial \mathbf{b}(\theta_i, r_i)}{\partial r_i}=-j\frac{4\pi\Delta{f}}{c}\mathbf{P_{\mathbf{A}}^{\perp}}[\mathbf{m}\odot\mathbf{b}(\theta_i, r_i)].\nonumber
\end{equation}
In the above two equations, $\mathbf{n}=[-(N-1)/2,\dots, (N-1)/2]^T$.

The CRB can be calculated by directly inverting the $\mathbf{J}_{\bm{\xi}}$ in (\ref{Eq:BlockFIM}). However, if the reflection amplitudes of different targets are statistically un-correlated, we can get a more explicit and intuitive expression of the CRB.

For targets with un-correlated amplitudes, if the snapshot number $L$ is sufficiently large,
\begin{equation}
\label{Eq:Uncorr}
[\mathbf{S}]_{ij}=\frac{1}{L}\sum_{l=0}^{L-1}a_i(l)a_j^*(l)\approx0, \mbox{ where } i\neq j.
\end{equation}
Then the off-diagonal blocks in (\ref{Eq:BlockFIM}) are zero, and the FIM is reduced to a diagonal block matrix. The inverse of $\mathbf{J}_{\bm{\xi}}$ is equal to the inversion of each diagonal block. In this case, the CRBs of  the $i$th  target's direction and range estimates are
\begin{equation}
\label{Eq:CRBDirection}
\mbox{CRB}_{\theta_i}=\frac{\sigma_n^2c^2}{2L[\mathbf{S}]_{ii}(4\pi f_cd\cos\theta)^2\gamma}|\mathbf{P}_{\mathbf{A}}^{\perp}(\mathbf{m}\odot\mathbf{b}(\theta_i, r_i))|^2,
\end{equation}
and
\begin{equation}
\label{Eq:CRBRange}
\mbox{CRB}_{r_i}=\frac{\sigma_n^2c^2}{2L[\mathbf{S}]_{ii}(4\pi\Delta{f})^2\gamma}|\mathbf{P}_{\mathbf{A}}^{\perp}(\mathbf{n}\odot\mathbf{b}(\theta_i, r_i))|^2,
\end{equation}
where
\begin{eqnarray}
\label{Eq:PerpCoef} 
\gamma&=&|\mathbf{P}_{\mathbf{A}}^{\perp}(\mathbf{m}\odot\mathbf{b}(\theta_i, r_i))|^2|\mathbf{P}_{\mathbf{A}}^{\perp}(\mathbf{n}\odot\mathbf{b}(\theta_i, r_i))|^2\nonumber\\
&&-\frac{1}{2}|(\mathbf{m}\odot\mathbf{b}(\theta_i, r_i))^H\mathbf{P}_{\mathbf{A}}^{\perp}(\mathbf{n}\odot\mathbf{b}(\theta_i, r_i))|^2\nonumber\\
&&-\frac{1}{2}\Re\left\{\left[(\mathbf{m}\odot\mathbf{b}(\theta_i, r_i))^H\mathbf{P}_{\mathbf{A}}^{\perp}(\mathbf{n}\odot\mathbf{b}(\theta_i, r_i))\right]^2\right\}.\nonumber\\
\end{eqnarray}

\textit{Remarks}: There are intuitive explanations of the results in (\ref{Eq:CRBDirection}-\ref{Eq:PerpCoef}). Due to the Cauchy-Schwartz inequality,
\begin{eqnarray}
\label{Eq:CS1}
&&|\mathbf{P}_{\mathbf{A}}^{\perp}(\mathbf{m}\odot\mathbf{b}(\theta_i, r_i))|^2|\mathbf{P}_{\mathbf{A}}^{\perp}(\mathbf{n}\odot\mathbf{b}(\theta_i, r_i))|^2\nonumber\\
&&\geq |(\mathbf{m}\odot\mathbf{b}(\theta_i, r_i))^H\mathbf{P}_{\mathbf{A}}^{\perp}(\mathbf{n}\odot\mathbf{b}(\theta_i, r_i))|^2,
\end{eqnarray}
and
\begin{eqnarray}
\label{Eq:CS2}
&&|(\mathbf{m}\odot\mathbf{b}(\theta_i, r_i))^H\mathbf{P}_{\mathbf{A}}^{\perp}(\mathbf{n}\odot\mathbf{b}(\theta_i, r_i))|^2\nonumber\\
&&\geq\Re\left\{\left[(\mathbf{m}\odot\mathbf{b}(\theta_i, r_i))^H\mathbf{P}_{\mathbf{A}}^{\perp}(\mathbf{n}\odot\mathbf{b}(\theta_i, r_i))\right]^2\right\}.
\end{eqnarray}
Thus $\gamma$ is no smaller than 0. Furthermore, the equalities in (\ref{Eq:CS1}) and (\ref{Eq:CS2}) will be valid when $\mathbf{m}=\mathbf{n}$. This result complies with direction-range coupling phenomenon in the LFDA. In the LFDA, the transmitting frequency shifts linearly among successive array elements, which means that $\mathbf{m}=\mathbf{n}$, $\gamma=0$,  and the CRBs approach infinity. However, in the RFDA, $\mathbf{m}$ is a random vector. Hence $\gamma\neq0$, and the CRBs are limited. This result explains in another perspective, why the estimation of direction and range in an RFDA are un-aliased.

\subsection{Mutual Coherence Based Performance Analysis}
\label{sub:Coherence}
As shown in Subsection \ref{Subsec:CS}, sparse recovery algorithms for compressive sensing are effective in target location and sidelobe elimination in the RFDA. In this subsection, a mutual coherence based guarantee will be provided for exact recovery in noiseless cases, and for the reconstruction error in noisy cases. 

In compressive sensing, the restricted isometry coefficient (RIC) is an effective metric to evaluate the recovery performance. However, finding the RIC of an observing matrix is exhaustive. In this paper, we adopt mutual coherence, which is much easier to compute, as an acceptable metric \cite{ben2010coherence} for deriving the recovery performance guarantee.

As defined in (\ref{Eq:CoherenceDef}), the calculation of mutual coherence is based on the inner products between the columns of the observing matrix. According to (\ref{Eq:SteeringVector}), these columns of the RFDA are determined by the directions and ranges of potential targets. However, it is quite difficult to give a closed form of mutual coherence for arbitrary directions or ranges. In this work, we will give the performance guarantee when potential targets are located at grid intersections, which we will call simply ``\textit{grids}".

\begin{definition}
\label{Df:Grids}
The \textit{grids} are directions and ranges satisfying the following two equalities:
\begin{equation}
S_a^N(\frac{2 f_cd\sin\theta}{c})=0\mbox{ or }N,\nonumber
\end{equation}
and
\begin{equation}
\Phi(\frac{2\Delta{f}r}{c})=0\mbox{ or }1.\nonumber
\end{equation}
\end{definition}

According to Definition \ref{Df:Grids}, there are $N$ direction grids from $-\pi/2$ to $\pi/2$, expressed as
\begin{equation}
\theta_k=\arcsin\frac{ck}{2f_cdN}, k=0,1,\dots, N-1.\nonumber
\end{equation}
However, the range grids depend on the distribution of $m_n$. For the discrete uniform distribution, there are $M$ direction grids from $0$ to $c\Delta{f}/2$, expressed as
\begin{equation}
r_i=\frac{c\Delta{f}i}{2M}, i=0,1,\dots, M-1.\nonumber
\end{equation}
Then there are $M\times N$ grids in an RFDA. Comparing (\ref{Eq:CoherenceDef}) with (\ref{Eq:BeamPatternDefinition}), the mutual coherence of the RFDA is equal to the the beampattern's highest sidelobe, where the directions and ranges are at grids.

With Theorem \ref{Th:AsyDistribution}, we can find that the real and imaginary parts of the beampattern are jointly Gaussian distributed. When it comes to mutual coherence, we have the following lemma.
\begin{lemma}
\label{LE:MutualCoherence}
If the carrier frequencies are discrete uniformly distributed, and all the potential targets are located at grids, the cumulative distribution of the mutual coherence of $\mathbf{\Phi}$ satisfies the following inequality:
\begin{equation}
\label{Eq:CdfCoherence}
\mbox{Pr}\{\mu<r\}\geq 1-(M-1)Ne^{-Nr^2}.
\end{equation}
\end{lemma}

\textbf{Proof:} With Definition \ref{Df:Grids} and Theorem \ref{Th:AsyDistribution}, it can be seen that the real and imaginary parts of sidelobes at grids (except where $\Phi(p)=1$, because according to (\ref{Eq:MeanBeamDef}) and (\ref{Eq:VarBeamPatternDef}), these sidelobes are deterministic and equal to zero) are zero-mean, \textit{i.i.d.} Gaussian. Their variances are
\begin{equation}
\sigma_r^2=\sigma_i^2=\frac{1}{2N}.
\end{equation}

Then the magnitudes of these sidelobes are Rayleigh distributed \cite{simon2007probability}, where
\begin{equation}
\label{Eq:Rayleigh}
\mbox{Pr}\{|\beta(q,p)|>r\}\leq e^{-Nr^2}.
\end{equation}

The number of grids where $\Phi(p)\neq1$ is $(M-1)N$, which means the probability that the highest sidelobe among the $(M-1)N$ grids is less than $r$ is
\begin{equation}
\label{Eq:PrMI}
\mbox{Pr}\{\mu < r\}\geq (1-e^{-Nr^2})^{(M-1)N},
\end{equation}
when $e^{-Nr^2}\ll1$. Furthermore, the right side of (\ref{Eq:PrMI}) can be approximated as
\begin{equation}
\label{Eq:PrApprox}
(1-e^{-Nr^2})^{(M-1)N}\approx 1-(M-1)Ne^{-Nr^2}.
\end{equation}
Substituting (\ref{Eq:PrApprox}) into (\ref{Eq:PrMI}), Lemma \ref{LE:MutualCoherence} is proven. \hfill{$\square$}

With the above lemma, we have the following guarantee for exact recovery in the noiseless case.
\begin{theorem}
\label{TH:Guarantee}
If the carrier frequencies are discrete uniformly distributed, and all the potential targets are located at grids, by using BP, the RFDA can exactly reconstruct $K$ targets  with a probability higher than $1-\epsilon$ ($\epsilon\leq1$) in the noiseless case, where
\begin{equation}
\label{Eq:Guarantee}
K\leq\frac{1}{2}\Big(1+\sqrt{\frac{N}{\ln(MN-N)-\ln\epsilon}}\Big).
\end{equation}
\end{theorem}

\textbf{Proof:} According to the corollaries in \cite{fuchs2004sparse}, if the mutual coherence
\begin{equation}
\mu\leq\frac{1}{2K-1},
\end{equation}
the $K$ non-zero entries of $\mathbf{x}$ can be exactly recovered in the noiseless case. Then with Lemma \ref{LE:MutualCoherence}, and by substituting $r=1/(2K-1)$ into (\ref{Eq:CdfCoherence}), we have that the probability of exact recovery is
\begin{equation}
\label{Eq:PrExactRecovery}
\mbox{Pr}(\hat{\mathbf{x}}=\mathbf{x})\geq 1-(M-1)Ne^{-N\left(\frac{1}{2K-1}\right)^2}.
\end{equation}
Thus the probability of exact recovery is larger than $1-\epsilon$, if
\begin{equation}
\label{Eq:ExactProbCondition}
1-(M-1)Ne^{-N\left(\frac{1}{2K-1}\right)^2} \geq 1-\epsilon.
\end{equation}

Theorem \ref{TH:Guarantee} is proven, because it is easy to verify that (\ref{Eq:ExactProbCondition}) is equivalent to (\ref{Eq:Guarantee}). \hfill{$\square$}

\begin{corollary}
\label{Co:ReconstructionError}
If the target number, $K$, satisfies  (\ref{Eq:Guarantee}), then in a noisy case, where the noise power of each array element is $\sigma_n^2$, the reconstruction error of QCBP satisfies the following constraint, with a probability higher than $1-\epsilon$:
\begin{equation}
\|\mathbf{x}-\hat{\mathbf{x}}\|_2\leq\frac{\sqrt{3(1+\eta)}}{1-(2K-1)\eta} (\sigma_l+\sigma_n),
\end{equation}
where $\eta=\sqrt{\left(\ln N+\ln(M-1)-\ln\epsilon\right)/N}$.
\end{corollary}

\textbf{Proof:} Corollary \ref{Co:ReconstructionError} can be proven by directly combining Theorem \ref{TH:Guarantee} in this paper and Theorem 2.1 in \cite{cai2010stable}. \hfill{$\square$}

Theorem \ref{TH:Guarantee} gives a sufficient condition for exact recovery in noiseless cases, and Corollary \ref{Co:ReconstructionError} gives an upper bound of reconstruction error in noisy cases. We noted that these two bounds are quite loose in practice, and are seeking tighter results.

\section{Numerical Results}
\label{sec:Results}

Numerical simulations verify the results achieved in this work, and demonstrate the merit of the new proposed FDA structure. In all the simulations, a linear array with $N=128$ elements was chosen as the archetype of the RFDA. The center carrier frequency $f_c$ was 3 GHz, and the frequency increment $\Delta{f} = 1$MHz. The inter-element distance $d=0.025$ m, which equals the quarter wavelength. For carrier frequency distributions that are discrete and continuous uniform, the parameter $M=64$. For Gaussian distributions, the standard deviation $\sigma=5$.

In the results, the expectations w.r.t. the random vector $\mathbf{m}$ are all calculated via 10,000 Monte-Carlo trials with different sample tracks of $\mathbf{m}$, unless otherwise specified. However, considering the massive computation time, in Subsection \ref{subsec:Results:detection} and Subsection \ref{subsec:Results:CRB} we conducted just 1,000 trials, with different noise but the same $\mathbf{m}$, to find the averaged successful detection rate and MSE for each SNR point.

With the above setup, simulations of the beampatterns' asymptotic distribution, target detection performance, and CRB/MSE of direction/range estimation were conducted. Results are provided in the following subsections.

\subsection{Asymptotic Distribution}
Because of the difficulties in direct verification of Theorem \ref{Th:AsyDistribution}, we decided to verify that the stochastic characteristics of $\rho(q,p)=e^{-j\alpha}\beta(q,p)$ match the results given in (\ref{Eq:MeanRho}), (\ref{Eq:VarRho}), and \ref{Eq:MeanSquare}). We also verified that the real and imaginary parts of $\rho(q,p)$ are Gaussian distributed.

The verifications of (\ref{Eq:MeanRho}) and (\ref{Eq:VarRho}) are similar to the simulations of the mean and variance of the beampattern. The results are omitted here because they are almost the same as those were shown in Fig. \ref{Fig:BeamMean} and Fig. \ref{Fig:BeamVar}.

\begin{figure}[!h]
\centering
\includegraphics[width=.5\textwidth]{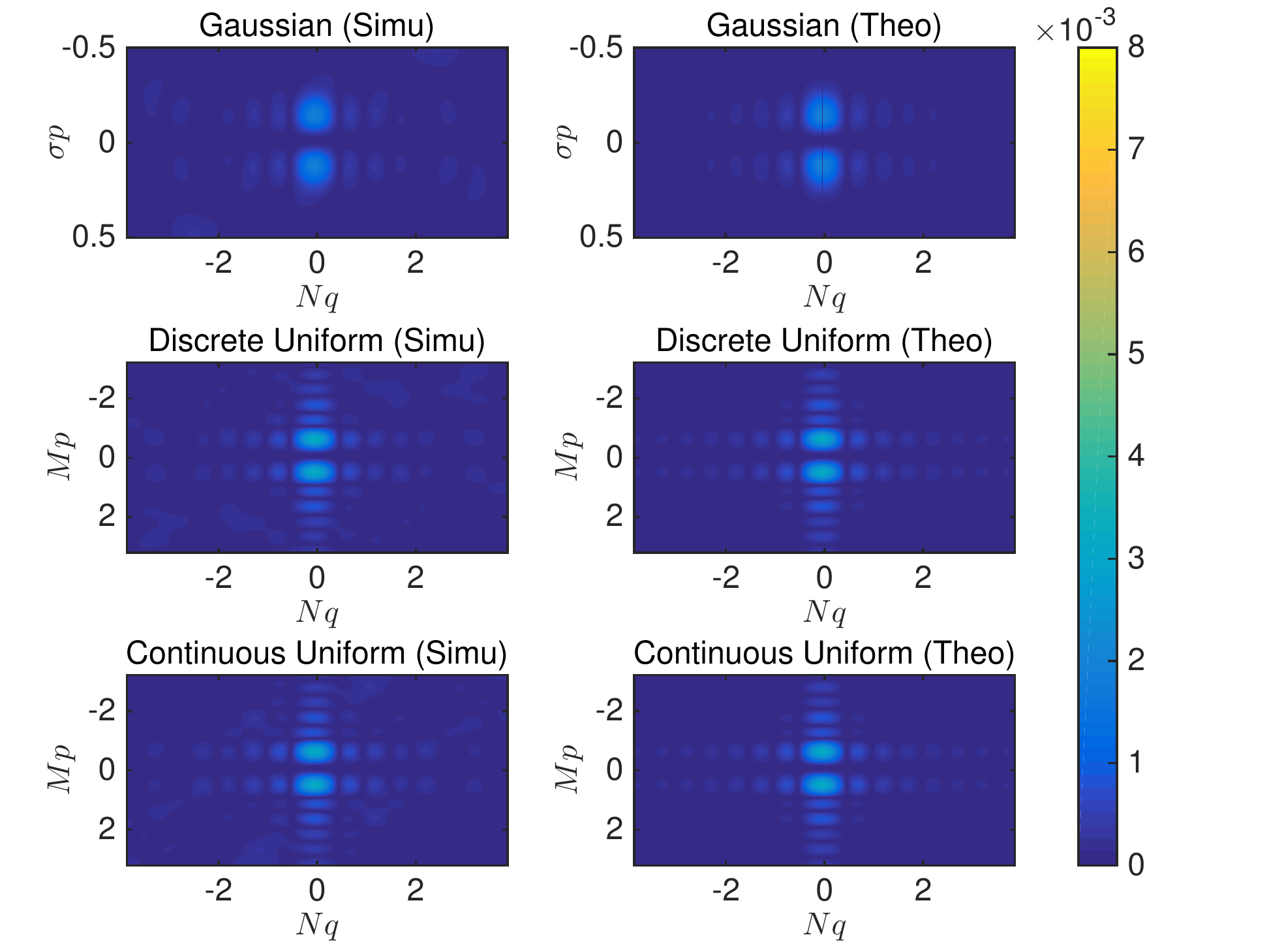}
\caption{The differences between the variances of the real and imaginary parts of the beampatterns. The right column shows the theoretical values provided in (\ref{Eq:MeanSquare}). The upper, middle, and lower rows are the results for Gaussian, discrete, and continuous uniformly distributed carrier frequencies, respectively.}
\label{Fig:PowerSub}
\end{figure}

For the verification of (\ref{Eq:MeanSquare}), we recorded the real and imaginary parts of $\rho(q,p)$ in each trial. The differences between the variance of $\Re\{\rho(q,p)\}$ and $\Im\{\rho(q,p)\}$ are shown in the left column of Fig. \ref{Fig:PowerSub}, and the theoretical values are displayed on the right. The simulated and analytical results match well. Moreover, it can be verified that the cross correlations between $\Re\{\rho(q,p)\}$ and $\Im\{\rho(q,p)\}$ for different $\{q, p\}$ pairs are very small (no larger that $10^{-4}$). Based on these facts, one can conclude that  (\ref{Eq:MeanSquare}) is verified.

Finally, we conducted the Kolmogorov-Smirnov (KS) test \cite{massey1951kolmogorov} for the normalized $\Re\{\rho(q,p)\}$ and $\Im\{\rho(q,p)\}$. When the trial number reached 10,000, all the $\{q, p\}$ pairs passed the test at a $5\%$ significance level.

With above results, the expression of asymptotic distribution of the beampattern in Theorem \ref{Th:AsyDistribution} can be considered as verified in this simulation setup.

\subsection{Target Detection Performance of Compressive Sensing}
\label{subsec:Results:detection}
Simulation results for the target detection performances of compressive sensing are provided in this subsection.

\begin{figure}[!h]
\centering
\includegraphics[width=.5\textwidth]{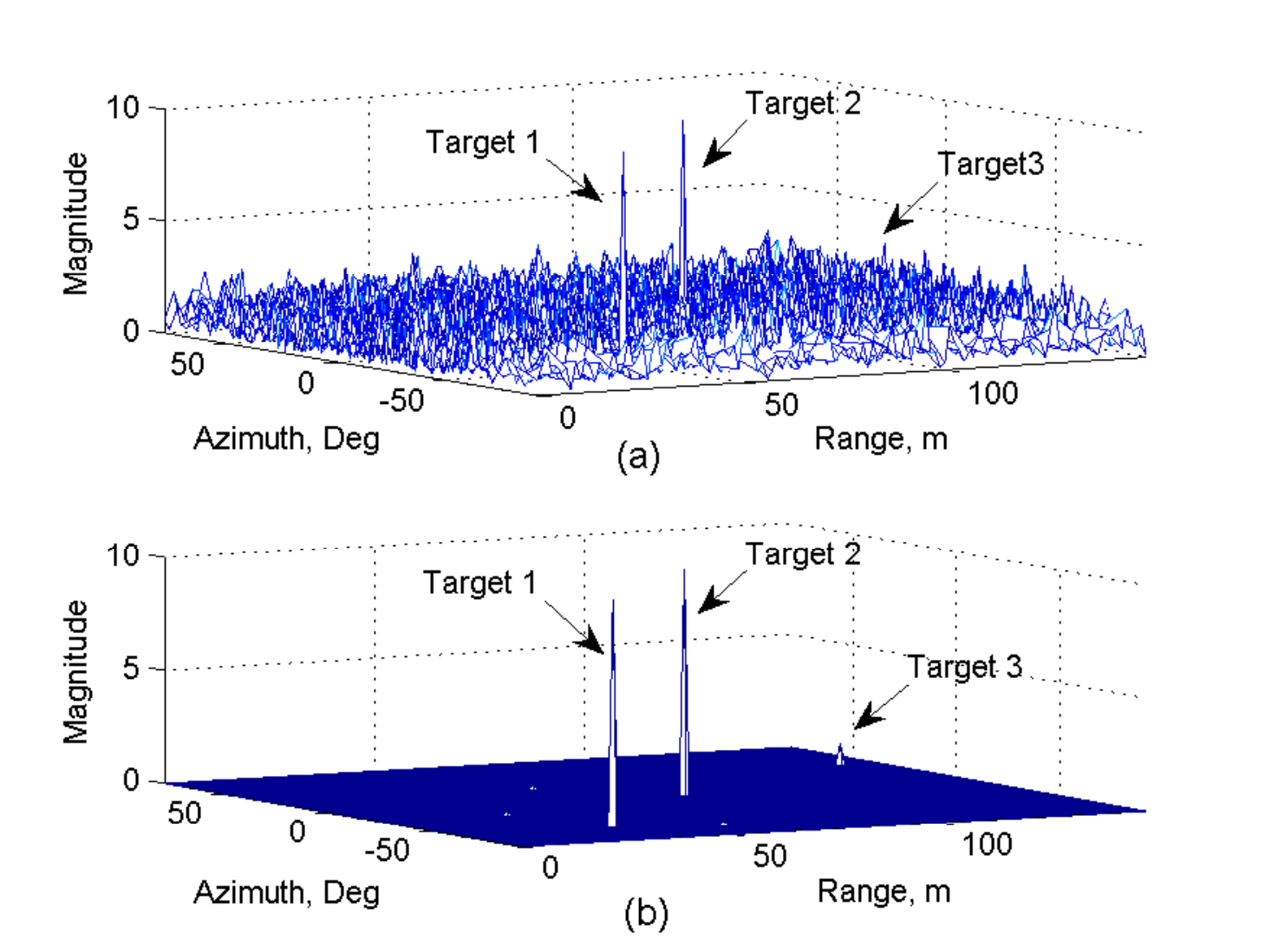}
\caption{Target direction and range location with an RFDA. (a) Beamforming result, (b) Compressive sensing result.}
\label{Fig:Beamforming}
\end{figure}

The first simulation gives an example of target indication. There were three targets at different ranges and directions ( $\theta_1=-30^o, r_1=10 \mbox{m}$;  $\theta_2=5^o, r_2=70 \mbox{m}$; $\theta_3=60^o, r_3=120 \mbox{m}$.). The amplitudes of Target 1 and Target 2 were identical and 10 dB larger than that of Target 3. The SNR of Target 3 was 0 dB (measured at the input of each receiver). Data was collected from only one snapshot. The beamforming result is shown in Fig. \ref{Fig:Beamforming}(a). The ranges and directions of Targets 1 and 2 are correctly indicated. But compared with the first two, Target 3 was too weak, and was masked by sidelobes. However, in the compressive sensing result (Fig. \ref{Fig:Beamforming}(b), by the SP algorithm), all the three targets were successfully detected, and their locations were correctly indicated as well.

\begin{figure}[!t]
\centering
\includegraphics[width=.5\textwidth]{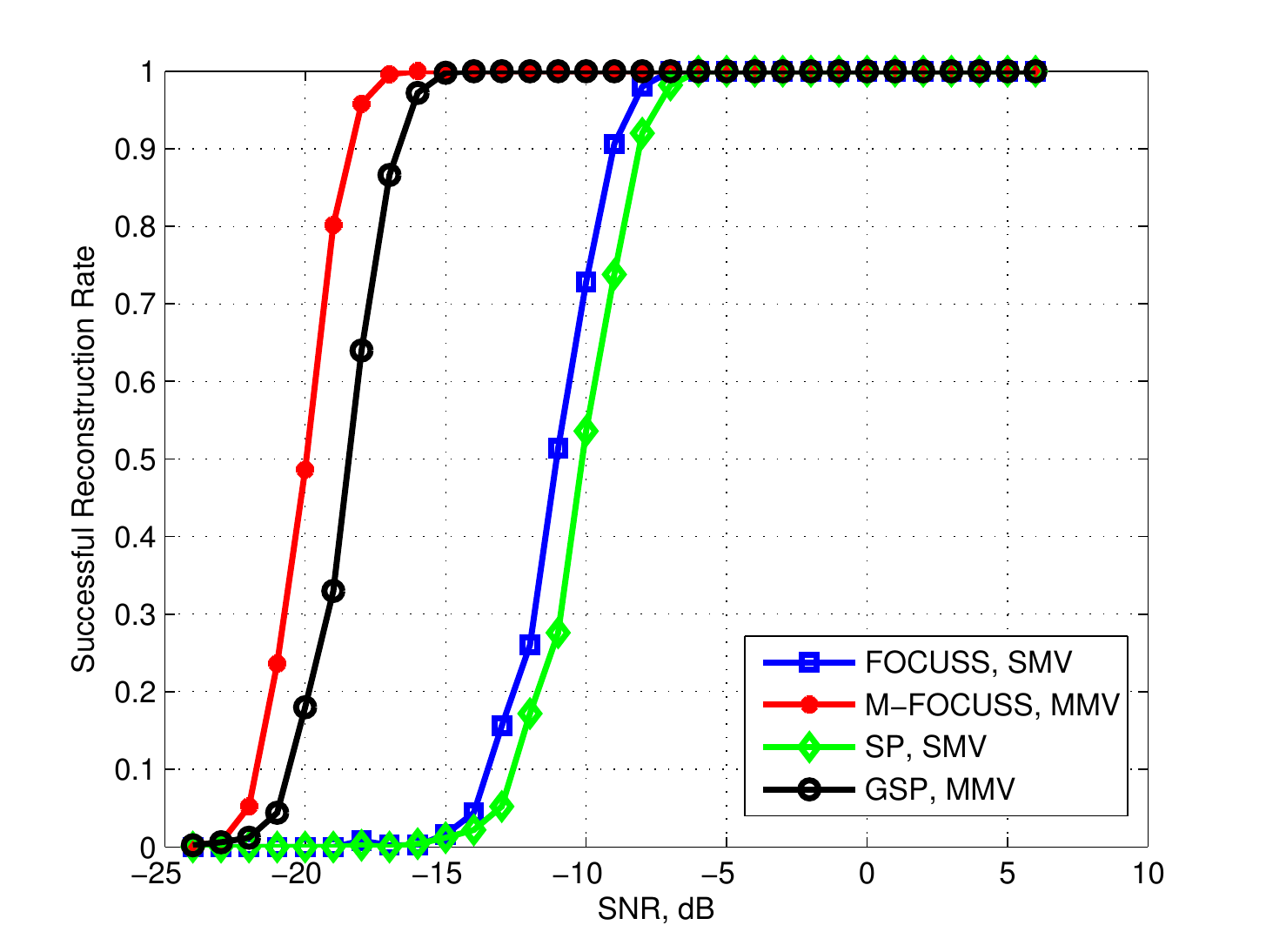}
\caption{Successful detection rates of different compressive sensing algorithms.}
\label{Fig:Pd}
\end{figure}

The second simulation was performed to evaluate the detection performances of different compressive sensing algorithms. There were two targets with identical reflection amplitudes, but different locations. The input SNR varied from -24 dB to 6 dB. A successful detection was defined as exact coincidence between the estimated and the true support sets. The results are shown in Fig. \ref{Fig:Pd}. In the comparison between the SMV and MMV scenarios, the detection performances of the MMV are better for both types of algorithms. In the comparison of recovery algorithm types, FOCUSS and M-FOCUSS outperform their subspace pursuit counterparts in both SMV and MMV scenarios.

\subsection{CRB and MSE of Direction/Range Estimation}
\label{subsec:Results:CRB}
\begin{figure}[!h]
\centering
\includegraphics[width=.5\textwidth]{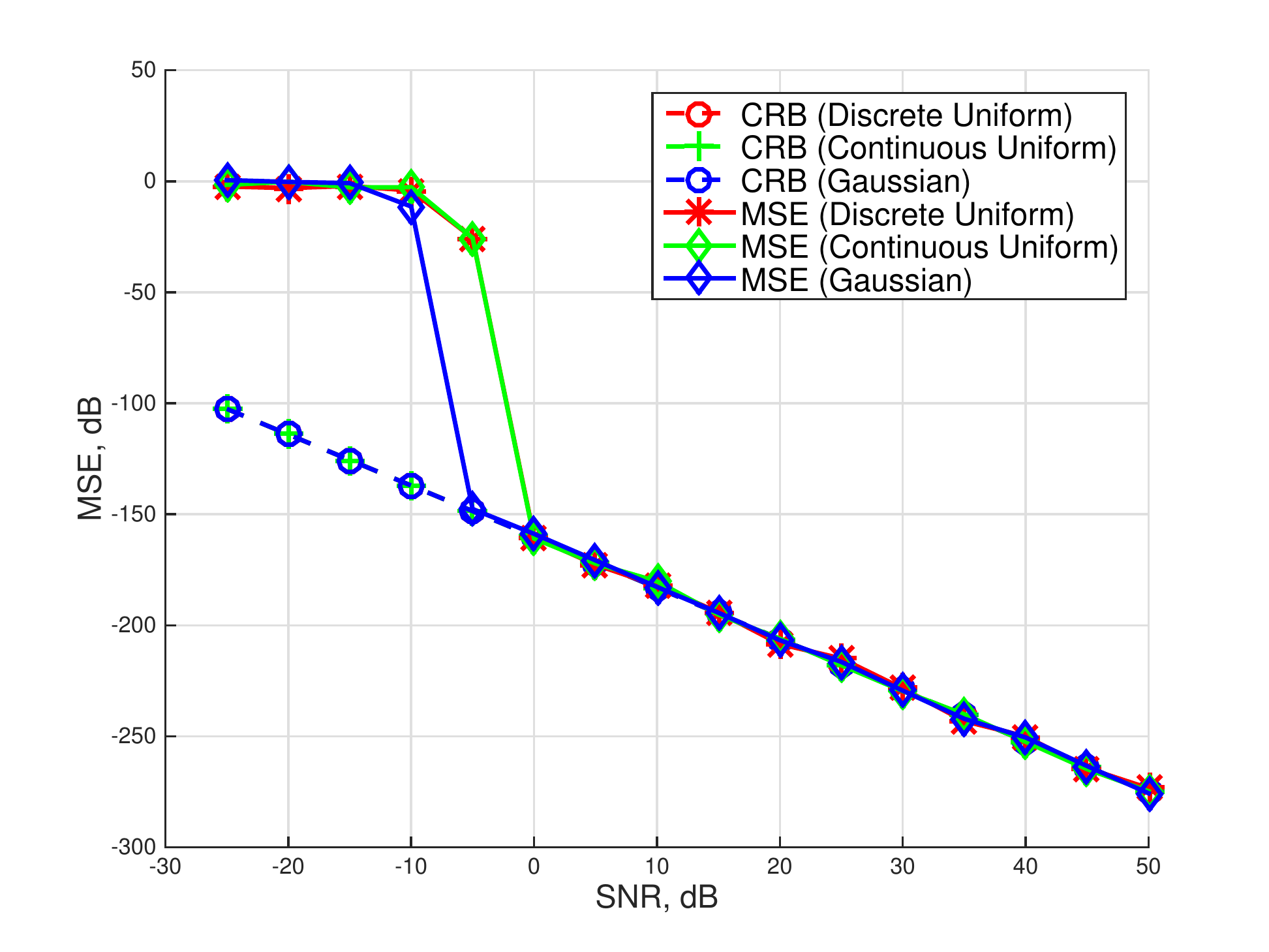}
\caption{MSEs of direction estimates obtained with ML estimator,
comparing with the corresponding CRBs given by (\ref{Eq:CRBDirection}).}
\label{Fig:CRB_direction}
\end{figure}

\begin{figure}[!h]
\centering
\includegraphics[width=.5\textwidth]{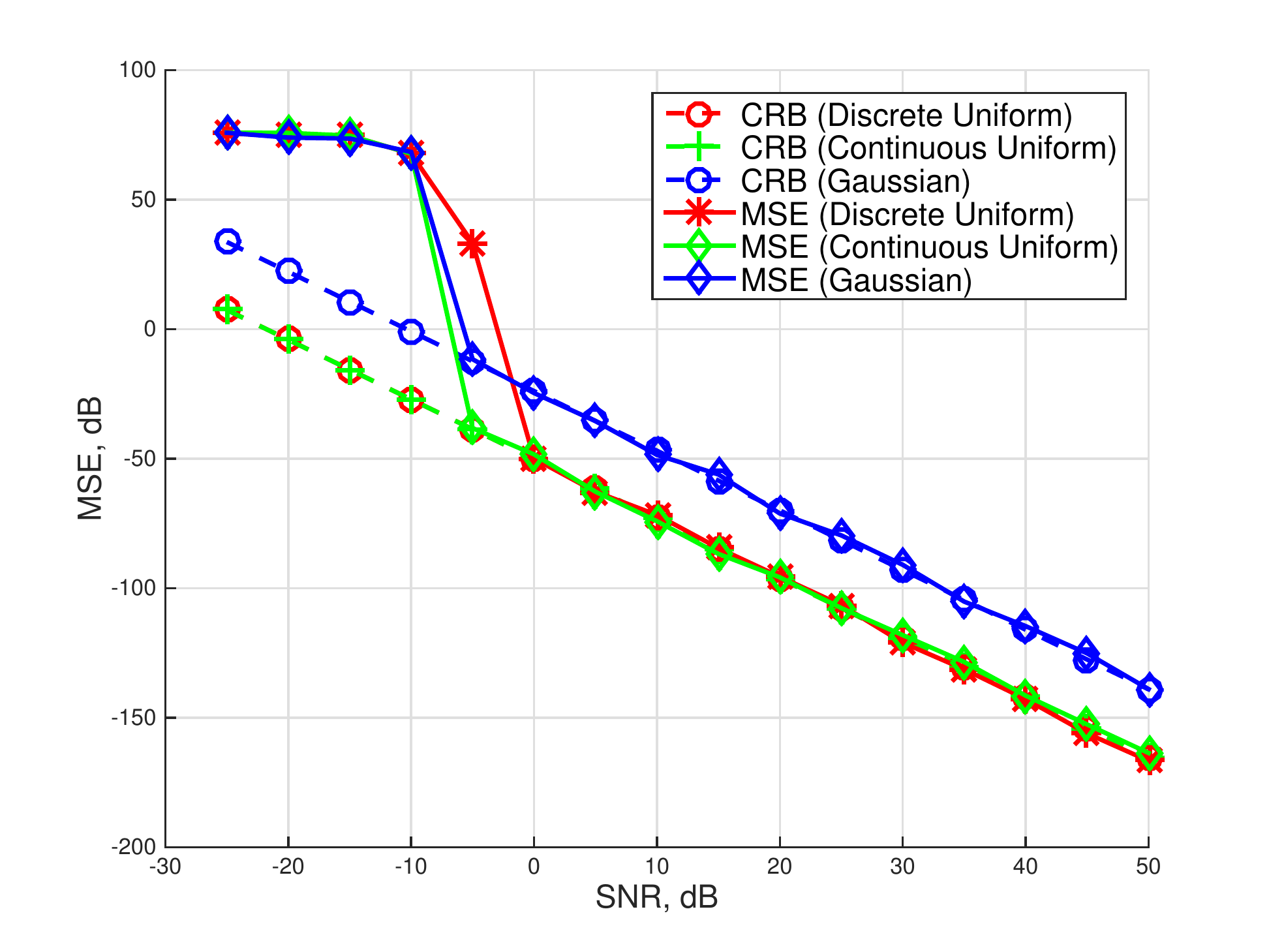}
\caption{MSEs of range estimates obtained with ML estimator,
comparing with the corresponding CRBs given by (\ref{Eq:CRBRange}).}
\label{Fig:CRB_r}
\end{figure}

In Fig. \ref{Fig:CRB_direction} and Fig. \ref{Fig:CRB_r}, the CRBs provided in (\ref{Eq:CRBDirection}) and (\ref{Eq:CRBRange}) are compared with corresponding MSEs of the direction and range estimates obtained via maximum likelihood (ML) estimation, for various SNRs. The amplitudes of the targets were kept constant, and noise power was varied to get each of the data points. A single target scenario was considered in this simulation.

The results are averaged w.r.t. different sample tracks of the random vector $\mathbf{m}$. It can be seen that the theoretical values of CRB match the MSEs well in high SNR situations (SNR $>0$dB). This result validates the correctness of the estimation error bounds provided in Subsection \ref{subsec:CRB}.

\section{Conclusion}
\label{sec:Conclusion}
We proposed a new frequency diverse array structure, named RFDA, to locate targets' directions and ranges without coupling. By randomly assigning the carrier frequencies of array elements, the RFDA realizes a random sparse sampling of the targets' information simultaneously in the spatial-frequency domain with a low system complexity. The beampattern of the RFDA is thumbtack-like, but with random sidelobe bases. Stochastic characteristics of the beampattern, the mean, variance, and asymptotic distribution were analytically derived, and two signal processing algorithms were introduced. In addition, the Cram\'{e}r-Rao bounds for direction/range estimation, as well as mutual coherence based limits for compressive sensing, were provided as performance guarantees of this new array. Numerical simulations demonstrated the RFDA's performance and verified the theoretical results.


\appendix
\label{AP:ProofofCov}
According to (\ref{Eq:VarYn}), the variance of $y_n(q,p)$ is independent of $n$, hence the sum of $\{y_n(q,p)\}_{n=0}^{N-1}$ is asymptotically complex Gaussian distributed \cite{kolmogorov1954limit}. Then the mean of $\bm{\beta}(q,p)$ is
\begin{equation}
\label{Eq:MeanRealImage}
\mathbb{E}_{\mathbf{m}}\left\{\left[\begin{array}{c} \beta_1(q,p)\\ \beta_2(q,p)\end{array}\right]\right\}
=\left[\begin{array}{c}\Re\{\bar{\beta}(q,p)\}\\ \Im\{\bar{\beta}(q,p)\}\end{array}\right].
\end{equation}
Substituting (\ref{Eq:MeanBeamDef}) into (\ref{Eq:MeanRealImage}), and noticing that $g(m_n)$ is even, we have
\begin{eqnarray}
\label{Eq:MeanReal}
\Re\{\bar{\beta}(q,p)\}&=&\frac{1}{N}S_a^N(q)\Phi(p)\cos{\alpha}\nonumber\\
\Im\{\bar{\beta}(q,p)\}&=&\frac{1}{N}S_a^N(q)\Phi(p)\sin{\alpha}.
\end{eqnarray}

The expression of $\mathbf{M}_{\bm{\beta}}$ can be derived as follows. Denote $z_n(q,p)=e^{-j\alpha}y_n(q,p)={1}/{N}\cdot e^{j2\pi\big(n-({N-1})/{2}\big)q}e^{j2\pi m_np}$, and then
\begin{eqnarray}
\label{Eq:DefofRho}
\rho(q,p)&\triangleq&\sum_{n=0}^{N-1}z_n(q,p)\nonumber\\
&=&e^{-j\alpha}\beta(q,p)\nonumber\\
&=&\frac{1}{N}\sum_{n=0}^{N-1}e^{j2\pi(n-\frac{N-1}{2})q}e^{j2\pi m_np}.
\end{eqnarray}
Moreover, the mean of $\rho(q,p)$ can be achieved directly by
\begin{equation}
\label{Eq:MeanRho}
\bar{\rho}(q,p)\triangleq \mathbb{E}_{\mathbf{m}}\left\{\rho(q,p)\right\}=\frac{1}{N}S_a^{N}(q)\Phi(p).
\end{equation}

By defining $\sigma_r^2$, $\sigma_i^2$ as the variances, and $\sigma_{ri}$ as the covariance of or between the real and imaginary parts of $\rho(q,p)$, we have that
\begin{equation}
\label{Eq:SumofVar}
\mathbb{E}_{\mathbf{m}}\left\{|\rho(q,p)-\bar{\rho}(q,p)|^2\right\}=\sigma_r^2+\sigma_i^2,
\end{equation}
and
\begin{equation}
\label{Eq:SqureofSub}
\mathbb{E}_{\mathbf{m}}\left\{[\rho(q,p)-\bar{\rho}(q,p)]^2\right\}=\sigma_r^2-\sigma_i^2+j\sigma_{ri}.
\end{equation}

Between $\rho(q,p)$ and $\beta(q,p)$, the only difference is the phase factor $e^{j\alpha}$, so they have same variances, given by
\begin{equation}
\label{Eq:VarRho}
\mathbb{E}_{\mathbf{m}}\left\{|\rho(q,p)-\bar{\rho}(q,p)|^2\right\}=\sigma^2_{\beta}(q,p)=\frac{1}{N}-\frac{1}{N}|\Phi(p)|^2.
\end{equation}

In addition, the explicit expression of $E\{[\rho(q,p)-\bar{\rho}(q,p)]^2\}$ can be derived by
\begin{eqnarray}
\label{Eq:MeanSquare}
&&\mathbb{E}_{\mathbf{m}}\left\{[\rho(q,p)-E\{\rho(q,p)\}]^2\right\}\nonumber\\
&=&\mathbb{E}_{\mathbf{m}}\left\{\rho^2(q,p)\right\}-\bar{\rho}^2(q,p)\nonumber\\
&=&\frac{1}{N^2}\mathbb{E}_{\mathbf{m}}\Big\{\sum_{l=0}^{N-1}\sum_{n=0}^{N-1}e^{j2\pi(n-\frac{N-1}{2})q}e^{j2\pi m_np}\nonumber\\
&&\cdot e^{j2\pi(l-\frac{N-1}{2})q}e^{j2\pi m_lp}\Big\}-\bar{\rho}^2(q,p)\nonumber\\
&=&-\bar{\rho}^2(q,p)+\frac{1}{N^2}\mathbb{E}_{\mathbf{m}}\Big\{\sum_{n=0}^{N-1}e^{j2\pi(n-\frac{N-1}{2})2q}e^{j2\pi m_n 2p}\Big\}\nonumber\\
&&+\frac{1}{N^2}\mathbb{E}_{\mathbf{m}}\Big\{\sum_{l=0, l\neq n}^{N-1}\sum_{n=0}^{N-1}e^{j2\pi(n+l-N+1)q}e^{j2\pi(m_n+m_l)p}\Big\}\nonumber\\
&=&-\bar{\rho}^2(q,p)+\frac{1}{N^2}S_a^N(2q)\Phi(2p)\nonumber\\
&&+\frac{1}{N^2}\Phi^2(p)\sum_{l=0, l\neq n}^{N-1}\sum_{n=0}^{N-1}e^{j2\pi(n+l-N+1)q}\nonumber\\
&=&\frac{1}{N^2}S_a^N(2q)\Phi(2p)-\frac{1}{N^2}\Phi^2(p)(S_a^N(q))^2\nonumber\\
&&+\frac{1}{N^2}\Phi^2(p)\left[(S_a^N(q))^2-S_a^N(2q)\right]\nonumber\\
&=&\frac{1}{N^2}S_a^{N}(2q)\cdot\left[\Phi(2p)-\Phi^2(p)\right].
\end{eqnarray}

Substituting (\ref{Eq:VarRho}-\ref{Eq:MeanSquare}) into (\ref{Eq:SumofVar}-\ref{Eq:SqureofSub}), and noting that $S_a^{N}(\cdot)$ and $\Phi(\cdot)$ are real, one finds that
\begin{equation}
\sigma_{ri}=0,
\end{equation}
\begin{equation}
\sigma_r^2=\frac{1}{2N}\left[1-\Phi^2(p)-\frac{S_a^N(2q)}{N}\left(\Phi^2(p)-\Phi(2p)\right)\right],
\end{equation}
and
\begin{equation}
\sigma_i^2=\frac{1}{2N}\left[1-\Phi^2(p)+\frac{S_a^N(2q)}{N}\left(\Phi^2(p)-\Phi(2p)\right)\right].
\end{equation}
Hence, the covariance matrix of the real and imaginary parts of $\rho(q,p)$ is
\begin{equation}
\mathbf{M}_{\mathbf{\rho}}(q,p)=\left[\begin{array}{cc}\sigma_r^2(q,p)&0\\0&\sigma_i^2(q,p)\end{array}\right].
\end{equation}
Because $\beta(q,p)=e^{j\alpha}\rho(q,p)$, we have that
\begin{eqnarray}
&&\mathbf{M}_{\bm{\beta}}(q,p)\nonumber\\
&&=\left[\begin{array}{cc}\cos\alpha&-\sin\alpha\\\sin\alpha&\cos\alpha\end{array}\right]\mathbf{M}_{\mathbf{\rho}}(q,p)\left[\begin{array}{cc}\cos\alpha&\sin\alpha\\-\sin\alpha&\cos\alpha\end{array}\right]\nonumber\\
&&=\left[\begin{array}{cc}\sigma_r^2\cos^2\alpha+\sigma_i^2\sin^2\alpha&\sin\alpha\cos\alpha(\sigma_r^2-\sigma_i^2)\\\sin\alpha\cos\alpha(\sigma_r^2-\sigma_i^2)&\sigma_r^2\sin^2\alpha+\sigma_i^2\cos^2\alpha\end{array}\right].\nonumber\\
\end{eqnarray}
Theorem \ref{Th:AsyDistribution} is proven. \hfill{$\square$}

\section*{Acknowledgment}
The authors would like to thank Mr. James Ballard for the proofreading.
\ifCLASSOPTIONcaptionsoff
  \newpage
\fi



\bibliographystyle{IEEEtran}
\bibliography{AWPL_CFDA.bbl}

\begin{thebibliography}{10}
\providecommand{\url}[1]{#1}
\csname url@samestyle\endcsname
\providecommand{\newblock}{\relax}
\providecommand{\bibinfo}[2]{#2}
\providecommand{\BIBentrySTDinterwordspacing}{\spaceskip=0pt\relax}
\providecommand{\BIBentryALTinterwordstretchfactor}{4}
\providecommand{\BIBentryALTinterwordspacing}{\spaceskip=\fontdimen2\font plus
\BIBentryALTinterwordstretchfactor\fontdimen3\font minus
  \fontdimen4\font\relax}
\providecommand{\BIBforeignlanguage}[2]{{%
\expandafter\ifx\csname l@#1\endcsname\relax
\typeout{** WARNING: IEEEtran.bst: No hyphenation pattern has been}%
\typeout{** loaded for the language `#1'. Using the pattern for}%
\typeout{** the default language instead.}%
\else
\language=\csname l@#1\endcsname
\fi
#2}}
\providecommand{\BIBdecl}{\relax}
\BIBdecl

\bibitem{yimin2016ICASSP}
Y.~Liu, ``Range azimuth indication using a random frequency frequency diverse
  array,'' in \emph{To be presented at the IEEE International Conference on
  Acoustics, Speech, and Signal Processing (ICASSP)}, 2016.

\bibitem{van2004detection}
H.~L. Van~Trees, \emph{Detection, Estimation, and Modulation Theory, Optimum
  Array Processing}.\hskip 1em plus 0.5em minus 0.4em\relax John Wiley \& Sons,
  2004.

\bibitem{merrill2001introduction}
I.~S. Merrill, \emph{Introduction to Radar Systems}.\hskip 1em plus 0.5em minus
  0.4em\relax McGrow-Hill, 2001.

\bibitem{Antonik2006Frequency}
P.~Antonik, M.~C. Wicks, H.~D. Griffiths, and C.~J. Baker, ``Frequency diverse
  array radars,'' \emph{IEEE Conference on Radar}, 2006.

\bibitem{wang2016overview}
W.-Q. Wang, ``Overview of frequency diverse array in radar and navigation
  applications,'' \emph{IET Radar, Sonar \& Navigation}, 2016.

\bibitem{xu2015space}
J.~Xu, S.~Zhu, and G.~Liao, ``Space-time-range adaptive processing for airborne
  radar systems,'' \emph{IEEE Sensors Journal}, vol.~15, no.~3, pp. 1602--1610,
  2015.

\bibitem{farooq2008exploiting}
J.~Farooq, M.~A. Temple, M.~Saville \emph{et~al.}, ``Exploiting frequency
  diverse array processing to improve {SAR} image resolution,'' in \emph{IEEE
  Radar Conference}.\hskip 1em plus 0.5em minus 0.4em\relax IEEE, 2008, pp.
  1--5.

\bibitem{xu2015range}
J.~Xu, S.~Zhu, and G.~Liao, ``Range ambiguous clutter suppression for airborne
  {FDA-STAP} radar,'' \emph{IEEE Journal of Selected Topics in Signal
  Processing}, vol.~9, no.~8, pp. 1620--1631, 2015.

\bibitem{higgins2009analysis}
T.~Higgins and S.~D. Blunt, ``Analysis of range-angle coupled beamforming with
  frequency-diverse chirps,'' in \emph{International Conference on Waveform
  Diversity and Design}.\hskip 1em plus 0.5em minus 0.4em\relax IEEE, 2009, pp.
  140--144.

\bibitem{huang2010frequency}
J.~Huang, ``Frequency diversity array: Theory and design,'' Ph.D. dissertation,
  UCL (University College London), 2010.

\bibitem{wang2014range}
W.-Q. Wang and H.~Shao, ``Range-angle localization of targets by a double-pulse
  frequency diverse array radar,'' \emph{IEEE Journal of Selected Topics in
  Signal Processing}, vol.~8, no.~1, pp. 106--114, 2014.

\bibitem{wangoptimal}
Y.~Wang, W.-Q. Wang, H.~Chen, and H.-Z. Shao, ``Optimal frequency diverse
  subarray design with {C}ram\'{e}r-{R}ao lower bound minimization,''
  \emph{IEEE Antennas and Wireless Propagation Lett.}, vol. 2014, pp.
  1188--1191, 2015.

\bibitem{wang2014transmit}
W.-Q. Wang and H.~C. So, ``Transmit subaperturing for range and angle
  estimation in frequency diverse array radar,'' \emph{IEEE Trans. on Signal
  Processing}, vol.~62, no.~8, pp. 2000--2011, 2014.

\bibitem{Li2008MIMO}
J.~Li and P.~Stoica, \emph{MIMO Radar Signal Processing}.\hskip 1em plus 0.5em
  minus 0.4em\relax Wiley-IEEE Press, 2008.

\bibitem{sammartino2013frequency}
P.~F. Sammartino, C.~J. Baker, and H.~D. Griffiths, ``Frequency diverse {MIMO}
  techniques for radar,'' \emph{IEEE Trans. on Aerospace and Electronic
  Systems}, vol.~49, no.~1, pp. 201--222, 2013.

\bibitem{ben2010coherence}
Z.~Ben-Haim, Y.~C. Eldar, and M.~Elad, ``Coherence-based performance guarantees
  for estimating a sparse vector under random noise,'' \emph{IEEE Trans. on
  Signal Processing}, vol.~58, no.~10, pp. 5030--5043, 2010.

\bibitem{lo1964mathematical}
Y.~T. Lo, ``A mathematical theory of antenna arrays with randomly spaced
  elements,'' \emph{IEEE Transactions on Antennas and Propagation}, vol.~12,
  no.~3, pp. 257--268, 1964.

\bibitem{kolmogorov1954limit}
A.~N. Kolmogorov and B.~Gnedenko, ``Limit distributions for sums of independent
  random variables,'' \emph{Addison-Wesley, Cambridge, Mass}, 1954.

\bibitem{simon2007probability}
M.~K. Simon, \emph{Probability Distributions Involving Gaussian Random
  Variables: A Handbook for Engineers and Scientists}.\hskip 1em plus 0.5em
  minus 0.4em\relax Springer Science \& Business Media, 2007.

\bibitem{rihaczek1969principles}
A.~W. Rihaczek, \emph{Principles of High-Resolution Radar}.\hskip 1em plus
  0.5em minus 0.4em\relax McGraw-Hill New York, 1969.

\bibitem{xu2015joint}
J.~Xu, G.~Liao, S.~Zhu, L.~Huang, and H.~C. So, ``Joint range and angle
  estimation using {MIMO} radar with frequency diverse array,'' \emph{IEEE
  Trans. on Signal Processing}, vol.~63, no.~13, pp. 3396--3410, 2015.

\bibitem{chen1998atomic}
S.~S. Chen, D.~L. Donoho, and M.~A. Saunders, ``Atomic decomposition by basis
  pursuit,'' \emph{SIAM Journal on Scientific Computing}, vol.~20, no.~1, pp.
  33--61, 1998.

\bibitem{malioutov2005sparse}
D.~Malioutov, M.~{\c{C}}etin, and A.~S. Willsky, ``A sparse signal
  reconstruction perspective for source localization with sensor arrays,''
  \emph{IEEE Trans. on Signal Processing}, vol.~53, no.~8, pp. 3010--3022,
  2005.

\bibitem{dai2009subspace}
W.~Dai and O.~Milenkovic, ``Subspace pursuit for compressive sensing signal
  reconstruction,'' \emph{IEEE Trans. on Information Theory}, vol.~55, no.~5,
  pp. 2230--2249, 2009.

\bibitem{gorodnitsky1997sparse}
I.~F. Gorodnitsky and B.~D. Rao, ``Sparse signal reconstruction from limited
  data using {FOCUSS}: A re-weighted minimum norm algorithm,'' \emph{IEEE
  Trans. on Signal Processing}, vol.~45, no.~3, pp. 600--616, 1997.

\bibitem{Feng2013Generalized}
J.~M. Feng and C.~H. Lee, ``Generalized subspace pursuit for signal recovery
  from multiple-measurement vectors,'' in \emph{IEEE Wireless Communications
  and Networking Conference (WCNC)}, 2013, pp. 2874 -- 2878.

\bibitem{cotter2005sparse}
S.~F. Cotter, B.~D. Rao, K.~Engan, K.~Kreutz-Delgado, and S.~Member, ``Sparse
  solutions to linear inverse problems with multiple measurement vectors,''
  \emph{IEEE Trans. on Signal Processing}, vol.~53, no.~7, pp. 2477--2488,
  2005.

\bibitem{kay1998fundamentals}
S.~Kay, \emph{Fundamentals of Statistical Signal Processing, Volume 1:
  Estimation Theory}.\hskip 1em plus 0.5em minus 0.4em\relax Prentice Hall PTR,
  1995.

\bibitem{stoica1990performance}
P.~Stoica and A.~Nehorai, ``Performance study of conditional and unconditional
  direction-of-arrival estimation,'' \emph{IEEE Trans. on Acoustics, Speech and
  Signal Processing}, vol.~38, no.~10, pp. 1783--1795, 1990.

\bibitem{fuchs2004sparse}
J.-J. Fuchs, ``On sparse representations in arbitrary redundant bases,''
  \emph{IEEE Trans. on Information Theory}, vol.~50, no.~6, pp. 1341--1344,
  2004.

\bibitem{cai2010stable}
T.~T. Cai, L.~Wang, and G.~Xu, ``Stable recovery of sparse signals and an
  oracle inequality,'' \emph{IEEE Trans. on Information Theory}, vol.~56,
  no.~7, pp. 3516--3522, 2010.

\bibitem{massey1951kolmogorov}
F.~J. Massey~Jr, ``The {Kolmogorov-Smirnov} test for goodness of fit,''
  \emph{Journal of the American Statistical Association}, vol.~46, no. 253, pp.
  68--78, 1951.

\end{thebibliography}

%
%

%

%
%
%




\end{document}